\documentclass[lettersize,journal]{IEEEtran}
\IEEEoverridecommandlockouts
\usepackage{comment}
\usepackage{algpseudocode}

\usepackage{cite}
\usepackage{amsmath,amssymb,amsfonts}
\usepackage{subcaption}
\usepackage{algorithm}
\usepackage{algpseudocode}
\usepackage{graphicx}
\usepackage{textcomp}
\usepackage{xcolor}
\def\BibTeX{{\rm B\kern-.05em{\sc i\kern-.025em b}\kern-.08em
    T\kern-.1667em\lower.7ex\hbox{E}\kern-.125emX}}
\usepackage{balance}
\usepackage{url}
\usepackage{booktabs} 
\usepackage{amsmath}  
\usepackage{bm}
\usepackage{lettrine}
\usepackage[hidelinks]{hyperref}
\usepackage{colortbl} 
\usepackage{xcolor} 
\usepackage{multirow}

\begin{document}
\title{Cyrus+: A DRL-based Puncturing Solution to URLLC/eMBB Multiplexing in O-RAN}

\author{Ehsan Ghoreishi,~\IEEEmembership{Graduate Student Member,~IEEE,} Bahman Abolhassani,~\IEEEmembership{Member,~IEEE,} \\
Yan Huang,~\IEEEmembership{Member,~IEEE,} Shiva Acharya,~\IEEEmembership{Graduate Student Member,~IEEE,} Wenjing Lou,~\IEEEmembership{Fellow,~IEEE,} Y.~Thomas Hou,~\IEEEmembership{Fellow,~IEEE}
\thanks{An abridged version of this paper appeared in the Proc. IEEE ICCCN, Hawaii, USA, July 29--31, 2024 \cite{Ghoreishi24:ICCCN:Cyrus}.}
\thanks{E. Ghoreishi, B. Abolhassani, S. Acharya, W. Lou, and Y.T. Hou are with Virginia Tech, Blacksburg, VA 24061.}
\thanks{Y. Huang is with NVIDIA Corp, Santa Clara, CA 95051.}
}


\maketitle

\begin{abstract}
Puncturing is a promising technique in 3GPP to multiplex Enhanced Mobile Broadband (eMBB) and Ultra-Reliable Low Latency Communications (URLLC) traffic on the same 5G New Radio (NR) air interface. 
The essence of puncturing is to transmit URLLC packets on demand upon their arrival, by preempting the radio resources (or subcarriers) that are already allocated to eMBB traffic.
Although it is considered most bandwidth efficient, puncturing URLLC data on eMBB can lead to degradation of eMBB’s performance.
Most of the state-of-the-art research addressing this problem employ raw eMBB data throughput as performance metric.
This is inadequate as, after puncturing, eMBB data may or may not be successfully decoded at its receiver.  
This paper presents Cyrus+---a deep reinforcement learning (DRL)-based puncturing solution that employs goodput (through feedback from a receiver's decoder), rather than estimated raw throughput, in its design of reward function.  
Further, Cyrus+ is tailored specifically for the Open RAN (O-RAN) architecture and fully leverages O-RAN's three control loops at different time scales in its design of DRL. 
In the Non-Real-Time (Non-RT) RAN Intelligent Controller (RIC), Cyrus+ initializes the policy network that will be used in the RT Open Distributed Unit (O-DU). 
In the Near-RT RIC, Cyrus+ refines the policy based on dynamic network conditions and feedback from the receivers.  
In the RT O-DU, Cyrus+ generates a puncturing codebook by considering all possible URLLC arrivals.  
We build a standard-compliant link-level 5G NR simulator to demonstrate the efficacy of Cyrus+.
Experimental results show that Cyrus+ outperforms benchmark puncturing algorithms and meets the stringent timing requirement in 5G NR (numerology 3).
 
\end{abstract}


\section{Introduction}\label{sec:Introduction}

Enhanced Mobile Broadband (eMBB) and Ultra-Reliable Low Latency Communications (URLLC) are two of the most important services that are supported by 5G New Radio (NR) \cite{3gpp:tr38300:NROveralDiscription}. 
eMBB aims to provide high data throughput for User Equipments (UEs) while URLLC aims to support latency-sensitive applications with high reliability 
requirements, with a transmission success probability of at least 99.999\% \cite{3gpp:tr38824:urllc}.
When eMBB and URLLC are transported on the same air interface, scheduling becomes a major challenge, due to their fundamentally different service requirements.

\begin{figure}[t]
    \centering
    \begin{subfigure}[b]{0.48\textwidth}
        \centering
        \includegraphics[width=1\linewidth]{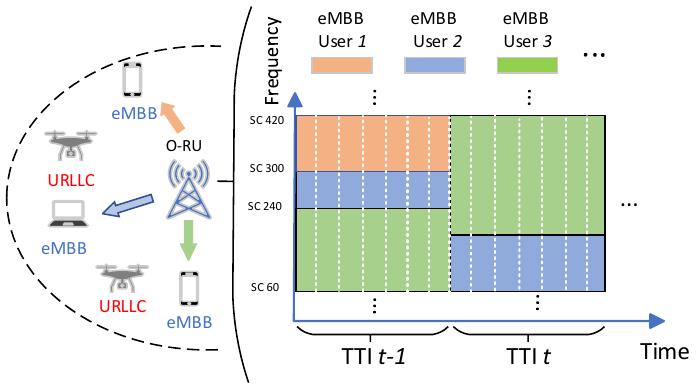}
        \caption{eMBB transmission only. No URLLC.}
        \label{fig:eMBB-URLLC_multiplexingsubfigure_a}
    \end{subfigure}
    \hfill
    \begin{subfigure}[b]{0.48\textwidth}
        \centering
        \includegraphics[width=1\linewidth]{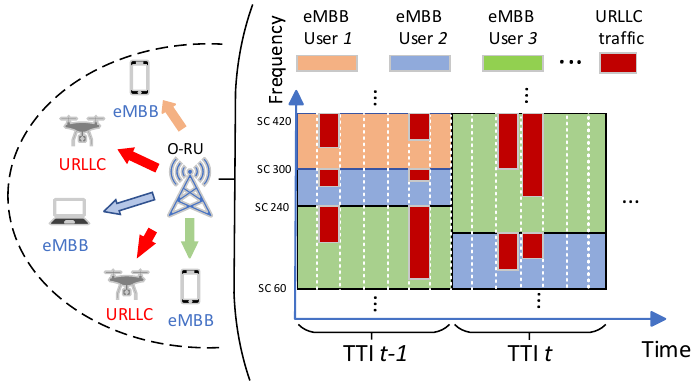}
        \caption{URLLC puncturing on eMBB.}
        \label{fig:eMBB-URLLC_multiplexingsubfigure_b}
    \end{subfigure}
    \caption{An illustration of multiplexing eMBB and URLLC on the same air interface through puncturing.}
    \label{fig:eMBB-URLLC_multiplexing}
\end{figure}

One promising approach proposed by the 3GPP is {\em puncturing\/} \cite{3gpp:tr38824:urllc, 3gpp:meeting88:Puncturing}. 
The essence of puncturing is that, in the absence of URLLC traffic, the eMBB users will enjoy all the radio resources; 
but when URLLC traffic arrives, it will be transmitted immediately by preempting (through ``puncturing" on) the radio resources that are already allocated to eMBB traffic.  
As shown in Fig.~\ref{fig:eMBB-URLLC_multiplexing}(a), the time axis is divided into transmission time intervals (TTI). 
Each TTI may consist of one or more time slots.\footnote{Without loss of generality, we assume a TTI consists of 1 time slot throughout this paper, as shown in Fig.~\ref{fig:eMBB-URLLC_multiplexing}.}
Each time slot is further divided into 2, 4, or 7 mini-slots \cite{3gpp:tr38824:urllc}.
Fig.~\ref{fig:eMBB-URLLC_multiplexing}(a) illustrates the case where one time slot is divided into 7 mini-slots, which is most ideal for supporting URLLC puncturing.  
In the frequency domain, resources are organized through subcarrier (SC) and SC spacing.
As shown in Fig.~\ref{fig:eMBB-URLLC_multiplexing}(a), the frequency axis is divided into SCs, where the SC spacing depends on 5G numerology (e.g., 120 kHz for numerology 3) \cite{3gpp:tr38211:PhysicalChannelsModulation}. 
In the absence of URLLC (Fig.~\ref{fig:eMBB-URLLC_multiplexing}(a)), the SCs on the channel is allocated to different eMBB users.
When URLLC traffic arrives, as depicted in red in Fig.~\ref{fig:eMBB-URLLC_multiplexing}(b), they are transmitted over the same air interface as eMBB by directly puncturing on scheduled eMBB traffic.

The immediate and indiscriminate puncturing of URLLC data on eMBB can lead to severe degradation of eMBB's performance (in terms of useful throughput, or goodput). 
So the fundamental question here is: {\em Is there a better (or even optimal) approach to perform URLLC puncturing on eMBB?\/}

A naive approach is to use eMBBs with the minimum number of SCs for URLLC puncturing (i.e., smallest eMBB first (SeF)). 
The idea is to limit puncturing exclusively to eMBB users with the minimum number of SCs and keep as many of the rest eMBBs intact as possible.
The problem with this approach is that it does not take advantage of the eMBB decoding capability at the receiver. 
In fact, at an eMBB receiver, it is possible to completely recover the original eMBB packet even though it has been punctured by URLLC. 
So a more effective strategy is to distribute (spread out) the URLLC's puncturing across multiple eMBB users for transmission.  
This strategy reduces the adverse impact on any single eMBB user and increases the probability that a punctured eMBB packet can still be recovered at its receiver.  
Unfortunately, there does not exist a fixed percentage or formula on how the eMBB packets should be punctured.
A resource proportional (RP) approach where the URLLC packets (or more precisely, their SCs) are punctured onto eMBB packets in portion to eMBB's packet size 
(see, e.g., \cite{Anand18:INFOCOM:Joint_URLLC_eMBB}) will not work well as it does not take into account the varying sensitivity (or immunity) among different eMBBs to puncturing.
So the question of how to optimally perform URLLC puncturing is far from trivial and remains an open problem.   

Unfortunately, it is not possible to formulate the URLLC puncturing problem in a closed form and make it suitable for traditional optimization techniques.
This is because the decoding behavior at a receiver side cannot be modeled mathematically, due to the complexity of physical (PHY) layer technologies (e.g., LDPC channel coding).
Further, when traffic behavior and network operating conditions are unknown and change over time, a model-based approach is futile in general.

To address these challenges, we present \emph{Cyrus+}\footnote{We use the codename ``Cyrus" after Cyrus the Great, who is known for his innovative governance of diverse cultures under a single empire, an attribute mirrored by our solution's ability to integrate eMBB and URLLC in the same air interface. 
The``+" sign denotes an improved version over our original design (Cyrus) in \cite{Ghoreishi24:ICCCN:Cyrus}.}---a model-free deep reinforcement learning (DRL) based URLLC data puncturing scheme. 
Architecturally, Cyrus+ is designed around the Open RAN (O-RAN) architecture \cite{O-RANAlliance2023:site:ORAN}, an open wireless ecosystem that promotes interoperability among different equipment vendors. 
Specifically, Cyrus+ leverages the three control loops in O-RAN design its algorithm vertically, in three different time scales.
In the Non-Real-Time (Non-RT) RAN intelligent controller (RIC), Cyrus+ initializes its puncturing policy network.
In Near-RT RIC, Cyrus+ learns and refines its policy continuously based on UEs' feedback.
In the open Distributed Unit (O-DU) and open Radio Unit (O-RU), Cyrus+ uses its policy network to find a feasible puncturing solution for URLLC in real-time. 

The main contributions of Cyrus+ are the following:

\begin{itemize}

\item Cyrus+ is the first URLLC puncturing solution specifically tailored for the O-RAN architecture.  
It is based on DRL and does not require any knowledge of assumptions on packet arrival, channel conditions, and decoding behavior.
At its core, Cyrus+ employs the {\em Soft Actor-Critic} (SAC) \cite{haarnoja18:ICML:SAC} method, which is particularly attractive due to its stability and sample efficiency.
It consists of one actor and two critics, with the actor in the 
RT O-DU and the two critics in the Near-RT RIC to meet the timing constraints.

\item 
In the Non-RT RIC, Cyrus+ sets the initial policy network for use in the RT O-DU. 
To address the convergence problem associated with a shallow NN (so as to meet RT requirement in O-DU), we employ an Automatic Curriculum Learning (ACL) method. 
This ACL allows the DRL agent to progressively learn and adapt its solution to the URLLC/eMBB puncturing problem.
With ACL, the DRL training remains stable and successfully converges as the agent gradually meets the most realistic operation conditions. 

\item 
In the Near-RT RIC, Cyrus+ aims to refine the policy based on changing network conditions. 
For the critic components (in SAC), we customize the NN parameter update mechanism to match our design for the RT O-DU component. 
We design a custom input buffer to store our data structure in the O-DU, effectively accommodating multiple actors and considering the differing time scales between TTI-based eMBB packets and mini-slot-based URLLC packets.

\item 
In the RT O-DU, the actor (in SAC) is meticulously designed to meet the stringent RT requirement of O-DU. 
We propose a parallel architecture to compute all possible URLLC arrivals concurrently to generate the puncturing codebook.
To ensure the policy network from each branch is feasible, 
we propose a feasibility enforcer for each branch that modifies the raw policy network to a feasible one should it violate the resource allocation constraint.

\item 
For validation, we build a system-level NR simulator for Cyrus+.
Within this simulator, we employ GPU to compute all branches in the O-DU in parallel, using Single Instruction Multiple Data (SIMD) for efficient matrix operations.
Using the NR simulator, we conduct a comprehensive evaluation of Cyrus+ and compare it to two benchmark puncturing schemes (RP and SeF). 
The results show that 
(i) 
Cyrus+ has good convergence behavior in learning; 
(ii) it consistently outperforms both RP and SeF in terms of the goodput under varying URLLC packet arrival, URLLC packet size, channel bandwidth and channel models; 
(iii) its O-DU meets the stringent timing requirement of NR numerology 3 (125 $\mu$s).  

\end{itemize}

\begin{table*}[ht]
\centering
\caption{Example prior efforts on URLLC puncturing based on the model-free approach.}
\label{table:Related_Work}
\begin{tabular}{|p{1.4 cm}|p{9 cm}|p{6 cm}|}
\hline
\textbf{References} & \textbf{Objectives and Constraints} & \textbf{Proposed Solutions} \\ \hline



\cite{Alqwider24:transaction:URLLCpuncturing_eMBBscheduling} & 
eMBB: Maximize eMBB's sum of data rate with minimum rate guarantee. \newline
URLLC: Minimize eMBB's sum of data rate loss under URLLC reliability and power constraints. & 
eMBB: a DRL (DDPG) agent \newline
URLLC: a DRL (DDPG) agent \\ \hline

\cite{Sohaib23:IEEEAccess:DRLapproximationDecoding} & 
Maximize the sum of eMBB raw throughput while satisfying URLLC reliability constraint. & 
For URLLC,  Double Deep Q-Network (DDQN)  \\ \hline


\cite{Fabio:GlobCom:DRL_threshold} & 
Minimize eMBB codewords outage while attempting to ensure URLLC latency stays below a threshold. & 
A DRL-based (PPO) agent \\ \hline

\cite{Huang20:IOT:DEMUX} & 
Maximize the sum PF utility of the eMBB UEs. & 
A DRL-based (DDPG) RB allocation for URLLC \\ \hline

\end{tabular}
\end{table*}

\section{Related Work} 
\label{sec:Related_Work}

There are some prior studies on optimizing URLLC and eMBB multiplexing on the same air interface. 
Basically, there are two approaches to multiplexing: reservation-based 
(see, e.g., \cite{3gpp:meeting88:Puncturing})
and puncturing 
(see, e.g., \cite{Anand18:INFOCOM:Joint_URLLC_eMBB, Bairagi:TC:Coexistence,  Alsenwi19:IEEEComLet:puncturing, Almekhlafi21:ICC:threshold, Almekhlafi21:transaction:puncturing_probability,   Alqwider24:transaction:URLLCpuncturing_eMBBscheduling, Sohaib23:IEEEAccess:DRLapproximationDecoding,  
Huang20:IOT:DEMUX,
Fabio:GlobCom:DRL_threshold}).  
Under a reservation-based approach, a block of SCs on the channel is reserved for URLLC in advance. 
Such an approach is most effective when URLLC traffic is heavy and predictable, so that advanced resource reservation for URLLC can ensure a clear separation between the eMBB and URLLC traffic while efficiently utilizing the channel bandwidth. 
However, when URLLC traffic is sparse and unpredictable (as in some sensing applications), a reservation-based scheme could be wasteful and ineffective.

A second approach to eMBB and URLLC multiplexing is puncturing, which is the focus of this paper. 
Under this approach, no channel bandwidth is reserved for URLLC traffic in advance. 
Instead, when URLLC packets arrive, they are punctured directly onto the eMBB packets. 
Such an approach could be effective when URLLC traffic arrival rate is low and the eMBB packets are capable of tolerating some puncturing errors.

Research under the puncturing approach can be classified into two branches: model-based 
(see, e.g., \cite{Anand18:INFOCOM:Joint_URLLC_eMBB, Bairagi:TC:Coexistence,  Alsenwi19:IEEEComLet:puncturing, Almekhlafi21:ICC:threshold, Almekhlafi21:transaction:puncturing_probability})
and model-free solutions 
(see, e.g., \cite{Sohaib23:IEEEAccess:DRLapproximationDecoding,  Huang20:IOT:DEMUX,  Alqwider24:transaction:URLLCpuncturing_eMBBscheduling, Fabio:GlobCom:DRL_threshold}).
Under model-based solutions, the problem is formulated using mathematical models and typically solved through analytical methods or optimization techniques.
Model-based solutions can be very effective when system behaviors and constraints can be accurately characterized and modeled in closed-form mathematical expressions.  
However, in our problem, the decoding behavior (to recover the adverse effect of puncturing), URLLC traffic arrival, and channel behavior are either mathematically intractable or unknown in advance.  
Any simplifying assumptions on these behaviors will degrade 
the performance of the final solutions. 
For example, in \cite{Anand18:INFOCOM:Joint_URLLC_eMBB, Alsenwi19:IEEEComLet:puncturing, Almekhlafi21:ICC:threshold}, the authors employed a threshold-based approach to model the effect of URLLC puncturing on eMBB packets (whether or not an eMBB packet can be successfully decoded) based on whether or not the number of puncturing URLLC SCs exceeds a threshold.
In \cite{Anand18:INFOCOM:Joint_URLLC_eMBB, Bairagi:TC:Coexistence, Huang20:IOT:DEMUX,
Almekhlafi21:transaction:puncturing_probability}, the authors assumed a linear relationship between the number of punctured resources from eMBB users and the achieved utility for eMBB users (e.g., in terms of achieved throughput).
Clearly, these assumptions greatly simplify mathematical modeling.
But their validity is difficult to justify.

In contrast to model-based solutions, model-free based solutions do not assume any predefined models or prior knowledge of system behavior to solve the problem.
Instead, they employ machine learning techniques (e.g., DRL) based on collected data.  
Table~\ref{table:Related_Work} summarizes some example 
model-free based solutions, highlighting their objectives, constraints, and learning approaches.
A common issue in some of these solutions (e.g.,  \cite{Sohaib23:IEEEAccess:DRLapproximationDecoding,   Alqwider24:transaction:URLLCpuncturing_eMBBscheduling}) is that when assessing the adverse impact of URLLC puncturing, there is no consideration of the actual decoding process at each eMBB receiver.  
In reality, all modern transmitters and receivers employ standard-based encoders and decoders. 
As such, the raw throughput estimated in these prior works is merely a lower bound for goodput after decoding.  
In other words, using raw throughput as the objective is not accurate to measure the impact of puncturing when the ultimate goal whether such raw throughput can be decoded successfully (i.e., goodput).
In \cite{Fabio:GlobCom:DRL_threshold}, the authors assumed that each URLLC packet has some delay tolerance that can be exploited for making puncturing decisions. 
Further, the authors employed inner erasure code for eMBB encoding/decoding and assumed knowledge of how many mini-slots an eMBB codeword can be punctured while can still recover. 
However, such an over-simplified encoding/decoding model for inner erasure code is not validated by a real inner erasure code in the simulation study. 
Further, modeling of radio resource is too coarse (on RB level rather than SC) and the impact of channel conditions on eMBB packet decoding was not considered.  
In \cite{Huang20:IOT:DEMUX}, the authors employed an actual decoder and conducted a reality check.
However, the authors assumed there is at most one URLLC packet per mini-slot, which unnecessarily restrictive. 
Further, the proposed solution in \cite{Huang20:IOT:DEMUX} is limited to support 5G NR numerologies 0 and 1.
It is also important to note that none of the prior works (i.e., \cite{Anand18:INFOCOM:Joint_URLLC_eMBB, Bairagi:TC:Coexistence,  Alsenwi19:IEEEComLet:puncturing, Almekhlafi21:ICC:threshold, Almekhlafi21:transaction:puncturing_probability,
Alqwider24:transaction:URLLCpuncturing_eMBBscheduling, Sohaib23:IEEEAccess:DRLapproximationDecoding,  
Huang20:IOT:DEMUX,
Fabio:GlobCom:DRL_threshold})
considered O-RAN in their design of learning-based puncturing solutions.

\begin{table}
    \centering
    \caption{Notation}
    \label{tab:symbols}
    \small
    \begin{tabular}{|p{1cm}|p{7cm}|}
        \hline
        Symbol & Definition \\
        \hline 
   
        \(\bm{b}_i(t)\) & The initial preemption vector for $i$ URLLC packet arrival in TTI $t$  \\
     \(\bm{b}^{*}_i(t)\) & A feasible preemption vector for $i$ URLLC packet arrival in TTI $t$ \\      
        \(\bm{B}^{*}(t)\) & The codebook matrix for TTI $t$  \\
        \(d_e(t)\) &  The eMBB decoding outcome at its receiver (success or failure) for TTI $t$\\ 
        \(E\) & Number of eMBB users within the coverage area \\
        \(\mathcal E\) & The set of eMBB users\\
        \(f_a\) & The policy neural network (actor) in O-DU \\
        \(f_0\) & The policy neural network (actor) in Non-RT RIC\\

        \(H\) & The number of samples from the input buffer randomly selected by each critic \\
        \(k(t, \tau) \) & The number of URLLC packets arriving during mini-slot $(t, \tau)$ \\ 
        \(\bm{k}(t) \) & The number of URLLC packets arriving in each mini-slot $\tau$ in a TTI $t$ \\ 
        \(L\) & The number of SCs needed to transmit a URLLC packet \\
        \(m_e(t, \tau)\) & The number of SCs on eMBB user $e$ being punctured during mini-slot $(t, \tau)$ \\      
        \(N\) & Total number of available SCs in a TTI\\        
        \(n_e(t)\) & The number of SCs allocated to eMBB user $e$ in TTI \(t\) \\
        \(p\) & Packet generation probability for a URLLC user in mini-slot $(t, \tau)$ \\  
        \(Q_1\) & The first Q-function estimator (critic 1) \\  
        \(Q_2\) & The second Q-function estimator (critic 2) \\
        \(r(t)\) & The reward function in TTI $t$ \\
        \( \bm{s}(t) \) & The scheduling vector for eMBB users in TTI $t$ \\
        \( \bm{S}(t) \)& The input matrix to the GPU in TTI $t$ \\
        \(t\) & The TTI index \\
        \(U\) & Number of URLLC users within the coverage area \\
        $\mathcal U$ & A set of URLLC users \\    
        \(\bm{y}(t, \tau)\) & The actual selected puncturing vector from the codebook for mini-slot $(t, \tau)$ \\     
        \(\bm{Y}(t)\) & The actual selected puncturing matrix 
        for all mini-slots in TTI $t$ \\
        \(\tilde{\bm{y}}(t, \tau)\) & The puncturing strategy sampled from the current policy during mini-slot $(t, \tau)$ \\  
        \(\zeta\) & The entropy regularization coefficient \\
        \(\pi_a^j\) & Probability distribution (stochastic policy $f_a$) 
        over actions for branch $j$ in O-DU \\
        \(\lambda\) & The reward discounted factor \\       
        \(\phi_i\) & The collection of trainable parameter in critic \(i\) \\    
        \hline
    \end{tabular}

\end{table}

\section{System Model and Problem Statement} 
\label{sec:Problem_Statement}

In this section, we formally describe the system behavior of URLLC puncturing on eMBB in the same 5G air interface and discuss the problem we aim to address. 
Table~\ref{tab:symbols} lists notations in this paper. 

\subsection{System Model}\label{subsec:System_Model}

Consider a 5G O-RAN small-cell in Fig.~\ref{fig:eMBB-URLLC_multiplexing} with an O-RU  transmitting to both eMBB and URLLC UEs in downlink direction.  
Denote $\mathcal{E}$ as a set of $E$ eMBB users and 
$\mathcal{U}$ as a set of $U$ URLLC users to which the O-RU transmit.
Referring to Fig.~\ref{fig:eMBB-URLLC_multiplexing}(a), time is divided into TTIs, with each TTI consisting of one {\em slot,\/} and each slot further divided into seven {\em mini-slots.\/} 
In the frequency domain, the available bandwidth is partitioned into subcarriers (SCs).
Denote $N$ as the total number of available SCs for the given bandwidth.

Denote $$\bm{s}(t) = [n_{1}(t), n_2(t), \cdots, n_{E}(t)]$$ as the scheduling vector for eMBB users in TTI $t$, where $n_{e}(t)$ denotes the number of SCs allocated to eMBB user $e$.
This eMBB vector $\bm{s}(t)$ is decided by the eMBB scheduler prior to TTI $t$ (e.g., during TTI $(t-1)$) and is assumed given {\em a priori\/} in this paper. 
Note that in each TTI $t$, some eMBB users $e \in {\mathcal E}$ may not be allocated with any SCs (either due to being inactive or there is simply not enough SCs). 
In this case, $n_e(t) = 0$.

For URLLC traffic at the O-RU, we assume each URLLC packet is of the same length and requires $L$ SCs in a mini-slot, 
$L < N$.  
For each URLLC UE $u \in {\mathcal U}$, we assume the probability that the O-RU will have a packet (in a mini-slot) destined to this URLLC UE follows a Bernoulli distribution with probability $p$.
To minimize the adverse impact of puncturing, the puncturing of URLLC's $L$ SCs may be distributed onto multiple eMBB packets.

Denote $k(t, \tau)$ as the number of URLLC packets that the BS generates for the URLLC users in mini-slot $(t, \tau)$. 
Then $k(t, \tau)$ follows a binomial distribution with parameters $(U, p)$. 
Given the channel bandwidth ($N$ SCs) and URLLC packet size ($L$ SCs), the maximum number of URLLC packets that can be punctured onto the eMBB traffic in a mini-slot is $\lfloor \frac{N}{L} \rfloor$ (corresponding to 
$\lfloor \frac{N}{L} \rfloor \times L$
SCs). 
This cap will ensure URLLC packet integrity and avoid fragmentation.  
Then, if $k(t, \tau)  > \lfloor \frac{N}{L} \rfloor$, the leftover  
$( k(t, \tau) - \lfloor \frac{N}{L} \rfloor)$ URLLC packets will need to be buffered at the O-RU and be transmitted in the next mini-slot.
Denote 
$$\bm{y}(t, \tau) = [m_1(t, \tau), m_2(t, \tau), \cdots, m_E(t, \tau)]$$ as the puncturing vector with respect to eMBB vector $\bm{s}(t)$ in mini-slot $(t, \tau)$),  
where $m_e(t, \tau)$ represents the number of eMBB SCs that are punctured by the URLLC packets and 
$0 \leq m_e(t, \tau) \leq n_e (t)$. 
Finding an optimal $\bm{y}(t, \tau)$ for each mini-slot is the focus of this paper.  

\subsection{Objective and Technical Challenges}\label{subsec:Objective}

\noindent \textbf{Objective} \ \  
Based on the above system model, it is evident that the performance of eMBB traffic is directly impacted by URLLC's puncturing behavior. 
Our objective is therefore to develop a puncturing mechanism for URLLC packets that minimizes their adverse impact on the eMBB data (i.e., eMBB packet decoding errors at the receiver), ensuring that the maximum amount of eMBB data can be successfully decoded at the receivers (i.e., goodput).

To measure the performance of a puncturing strategy, it is necessary to incorporate the eMBB decoding outcome at its receiver (success or failure) into the metric.  
For this purpose, we denote $d_e(t)$ as a binary indicator function with the following definition: 
\[
{d_e(t)} = \begin{cases} 
1 & \text{if the eMBB user $e$'s packet in TTI $t$ can be} \\ & \text{ decoded successfully,} \\
0 & \text{otherwise}.
\end{cases}
\]
Then the goodput in TTI $t$ is 
$\sum_{e \in \mathcal{E}} d_e(t) \cdot n_e(t)$ and the average goodput over a window $T$ is 
$\frac{1}{T} \sum_{t=1}^T\left[ \sum_{e \in \mathcal{E}} d_e(t) \cdot n_e(t) \right]$. 
So our objective is to maximize this long term average goodput through optimal puncturing, i.e., 
\begin{equation}
\label{eq:costmin_t}
\begin{split}
    &\max _{\bm{y}(t, \tau)} \lim _{T \rightarrow \infty} \frac{1}{T} \sum_{t=1}^T\left[ \sum_{e \in \mathcal{E}} d_e(t)n_e(t) \right]
\end{split}
\end{equation}
where $\bm{y}(t, \tau)$ is the decision vector for URLLC puncturing.

\noindent \textbf{Technical Challenges} \ \  
Although we are able to characterize our objective function (\ref{eq:costmin_t}) in closed form, it is, however, not possible to develop a tractable model for the complete problem (e.g., constraints). 
The major challenge lies in the difficulty in characterizing decoding behavior of eMBB packets, i.e., whether or not a punctured eMBB packet can be decoded successfully at its receiver. 
This decoding behavior cannot be modeled accurately due to the complexity in the 5G NR PHY layer, particularly with LDPC coding \cite{Gallager1963:IEEETrans:LDPC}. 
LDPC is the primary forward error correction (FEC) mechanism for 5G NR data transmission \cite{3gpp23:tr38201:PHY, Gallager1963:IEEETrans:LDPC}. 
However, the decoding process for LDPC is inherently complex, making it impossible to use a closed-form expression for its behavior.\footnote{For a lower bound on LDPC's error-correction capability, albeit not in closed form, refer to \cite{Rybin14:IEEE:errorcorection}.}
Consequently, we are unable to derive an explicit expression for the BLER performance of eMBB. 
BLER performance can only be determined through experimental or simulation-based methods \cite{Richardson18:BLER:Numerically}.

Given the absence of a closed-form model for the decoding behavior, one might consider using a simplified approximation model (see, e.g., \cite{ Anand18:INFOCOM:Joint_URLLC_eMBB, Alsenwi19:IEEEComLet:puncturing, Almekhlafi21:ICC:threshold, Bairagi:TC:Coexistence}). 
However, these simplified models suffer from poor performance in practice as they cannot capture the intricate effects of URLLC puncturing on eMBB performance, as we shall show in Section~\ref{sec:Performance}.

Given the aforementioned challenge in establishing a model-based formulation to the URLLC puncturing problem, we propose a learning-based approach in this paper, predicated on the O-RAN reference architecture.

\section{Cyrus+: An Overview} 
\label{sec:Cyrus+} 

\subsection{Basic Idea} 
\label{subsec:basic_idea} 

While the number of URLLC packets arriving in each mini-slot $k(t, \tau)$ may vary widely, the maximum number of URLLC packets that can be served in a mini-slot is capped at $\lfloor \frac{N}{L} \rfloor$ (see our discussions in Section~\ref{subsec:System_Model}). 
In other words, for each mini-slot, the number of URLLC packets that can puncture eMBB packets is $\min \left( \left\lfloor \frac{N}{L} \right\rfloor, k(t, \tau) \right)$ and can only be a number from \(0, 1, \cdots, \lfloor \frac{N}{L} \rfloor\).
This suggests that for the given eMBB allocation $\bm{s}(t)$, if we can pre-compute the puncturing strategies for all 
possible URLLC packet arrival scenarios 
in a mini-slot ($\lfloor \frac{N}{L} \rfloor + 1$ total)  before hand (assuming all channel remains coherent for all mini-slots in a TTI), then we can simply pick up the optimal puncturing strategy upon the arrival of the actual number of URLLC packets in a mini-slot.   
That is, we can pursue a proactive approach by simply reading from a pre-computed 
``codebook" (or ``puncture-book").

\begin{table*}[t]
\centering
\renewcommand{\arraystretch}{1.2} 
\caption{Comparison of different model-free DRL algorithms.}
\label{table:Comparison}
\begin{tabular}{|l|l|l|l|}
\hline
\textbf{Model-free DRL}  & \multicolumn{1}{l|}{\textbf{Key Features}} & \textbf{Convergence} & \textbf{Sample Efficiency} \\ \hline
PPO \cite{Schulman:arXiv:PPO, Haijun20:SpectrumSharing:PPO}  &
\begin{tabular}[l]{@{}l@{}} 
On-policy, policy gradient-based, supports discrete and continuous\\  actions, most simple to implement\end{tabular}  
& Slow but stable 
& Low \\ \hline
DDPG \cite{Timothy19:arXiv:DDPG, Xuemin20:DDPG:resourceManagment} & \begin{tabular}[l]{@{}l@{}} Deterministic policy, continuous action spaces, actor-critic \\  framework, add noise to increase exploration \end{tabular} & 
Slow and unstable   &
Moderate  \\ \hline
TD3 \cite{Fujimoto18:Arkiv:TD3} &
\begin{tabular}[l]{@{}l@{}} Deterministic policy, twin critics, delayed actor updates, reduces\\  overestimation bias, improvement over DDPG \end{tabular} 
& Fast and stable 
& High   \\ \hline
SAC \cite{haarnoja18:ICML:SAC}  & \begin{tabular}[l]{@{}l@{}} Stochastic policy, entropy maximization, twin critics, delayed actor\\  updates, reduces overestimation bias, improvement over DDPG\end{tabular} & Fast and stable 
& High  \\ \hline
\end{tabular}
\end{table*}


To pre-compute this puncture-book, we propose to develop a modified version of the SAC algorithm \cite{haarnoja18:ICML:SAC}. 
SAC's strength lies in its ability to continuously learn and improve pre-computed policies.
For our problem, the modified SAC algorithm consists of two key components: actor and critics.
The actor is responsible for making decisions, i.e.,  determining the optimal puncturing strategies for each of the $(\lfloor \frac{N}{L} \rfloor +1)$
possible URLLC packet arrival scenarios. 
Meanwhile, the critics provide continuous feedback to the actor, helping to refine and improve the decision-making process by adapting to changing environmental conditions over time.

To meet the stringent real-time (mini-slot) requirements of URLLC, 
Cyrus+ employs COTS GPU to speed up the actor's decision-making process.
The goal is to ensure its computation time be within the duration of a TTI, so that the puncture-book is readily available at the start of each TTI (to be used for each mini-slot in the TTI).

\subsection{Why SAC?}
\label{subsec:Why_SAC}
Due to the unknown nature in the puncturing problem (e.g., random URLLC packet arrival, eMBB packet payload, dynamic channel conditions), it is impossible to rely on a static scheduling policy. 
Instead, it is necessary to continuously learn from the environment to update the policy, allowing the scheduler to dynamically adjust to the time-varying parameters. 
DRL can address this need, where the agent interacts with the environment in real time and, based on the feedback from these interactions, iteratively improves its policy.

In the domain of the DRL, there are two approaches: 
model-based and model-free \cite{Fabio:GlobCom:DRL_threshold}.
Model-based DRL explicitly attempts to learn or estimate the environment's dynamics by approximating the transition probability function and reward function \cite{sutton2018:Book:RL:AnIntroduction}. 
This approach is effective in environments where system dynamics, such as how actions influence future states and outcomes, are well understood. 
However, this approach presents challenges in our URLLC puncturing problem as the 5G PHY layer is too intricate to be captured by a closed-form expression (as discussed in \ref{subsec:Objective}).
In contrast, model-free DRL does not attempt to model the environment explicitly. 
Instead, it focuses on learning the optimal policy or value function directly from experience, through interactions with the environment \cite{sutton2018:Book:RL:AnIntroduction}, which is more suitable to address our problem.

There are various model-free DRL algorithms, such as Proximal Policy Optimization (PPO) \cite{Schulman:arXiv:PPO, Haijun20:SpectrumSharing:PPO}, Deep Deterministic Policy Gradient (DDPG) \cite{Timothy19:arXiv:DDPG, Xuemin20:DDPG:resourceManagment}, Twin Delayed DDPG (TD3) \cite{Fujimoto18:Arkiv:TD3}, and SAC \cite{haarnoja18:ICML:SAC}. 
Table~\ref{table:Comparison} presents a comparison of these different model-free algorithms.
Among them, SAC emerges as the most suitable choice for our puncturing problem.
SAC's primary advantage lies in its dual objectives: maximizing both expected rewards and policy entropy. 
By directly optimizing entropy through its loss function, SAC encourages randomness in action selection, prompting the agent to explore a broader range of actions rather than prematurely settling on a predictable strategy.
In contrast, DDPG and TD3 use deterministic policies, which select the same action for a given state during execution. 
Although deterministic policies are straightforward to implement, their lack of flexibility can lead to suboptimal performance in rapidly changing conditions, as they fail to adequately explore alternative strategies once the policy is learned.
DDPG and TD3 tend to struggle in unpredictable environments, especially when the action space is large (e.g., $\binom{L+E-1}{E-1}$ in our case), and dynamic adaptation is necessary.

Moreover, SAC incorporates twin critics in its structure, with delayed policy updates.
The two critics operate independently, with each critic estimating the Q-value for a given state-action pair. 
The actor then updates its policy based on the minimum of these two Q-values. 
This mechanism reduces overestimation bias, which enhances stability compared to DDPG and PPO,
providing more accurate value estimates and accelerating the learning process.

Finally, SAC is an off-policy algorithm, allowing it to reuse past experiences during training, which significantly enhances sample efficiency compared to on-policy algorithms.
By reusing past experiences, we mean SAC can replay old data, allowing it to make more updates to its policy with fewer interactions with the environment.
In contrast, on-policy methods, such as PPO, require collecting new data for each training iteration, making it less sample-efficient.

Despite SAC's strengths, applying it to solve our URLLC puncturing problem is not without challenges. 
One primary difficulty is that while the SAC algorithm excels in environment with continuous action spaces, its adaptation to high-dimensional discrete action spaces introduces significant complexities.
This adaptation complicates both the formulation of the policy and the learning process, as it requires bridging the gap between the continuous outputs from the policy network and the discrete action requirements. 
These changes can impair the agent's ability to explore effectively and converge on optimal solutions, potentially increasing the computational load and complicating the tuning of hyperparameters.
Moreover, SAC’s dependency on a wide range of hyperparameters, such as learning rates and reward scaling, can make its deployment challenging.
We will address these issues in our design of Cyrus+.  
 
\begin{figure}[t]
    \centering
    \includegraphics[width=\linewidth]{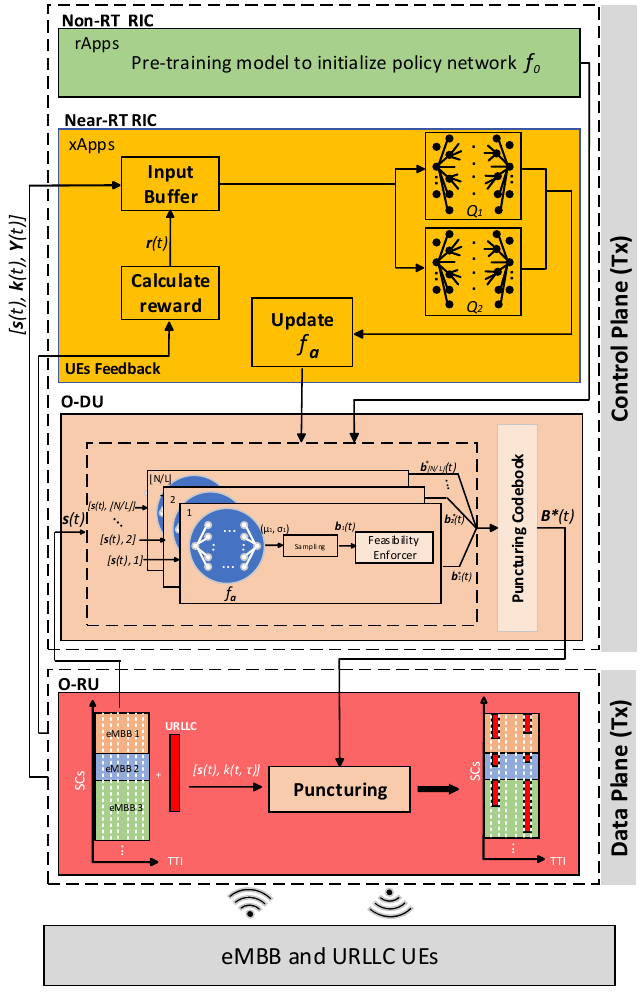} 
    \caption{An overview of Cyrus+ in the O-RAN architecture.}
    \label{fig:ORANDiagram}
\end{figure}

\subsection{Cyrus+: An Overview}
\label{subsec:Cyrus_Overview} 

Figure~\ref{fig:ORANDiagram} illustrates architecture 
of Cyrus+, which sits on the control plane, with a focus on 
producing a puncturing codebook for each TTI. 
On the data plane, the puncturing codebook is applied to puncture  eMBB packets on the mini-slot level to transport URLLC packets.
In the rest of this section, we give an overview both planes.

\subsubsection{Control Plane}

Our proposed Cyrus+ in the control plane is customized for the O-RAN architecture \cite{O-RANAlliance2023:site:ORAN} through its Non-RT RIC, Near-RT RIC, and O-DU.

\noindent \textbf{Non-RT RIC}\ \
In this block, the objective is to pre-train a neural network (NN) (denoted as $f_0$) so that we can have an initial policy to start with. 
This process entails establishing suitable values for the weights in the NN.
The training data includes historical scheduling configurations for eMBB users and the possible number of URLLC packets that can be accommodated in each mini-slot.

Once the initial policy network $f_0$ is trained, it is transferred to the Near-RT RIC via A1 interface, and then to RT O-DU via E2 interface.  
The O-DU then utilizes this initial policy to compute feasible puncturing vectors (so as to build a puncturing codebook) in one TTI.  

\noindent \textbf{Near-RT RIC}\ \
The goal of the xApps in the Near-RT RIC is to periodically update the current policy network, denoted as \(f_a\) (initialized with \(f_0\)).
Within this component, the critics (\(Q_1\) and \(Q_2\)) of the SAC algorithm are employed.
These critics evaluate the efficacy of the puncturing codebook executed in the data plane.
This evaluation is facilitated by calculating a reward function (based on UEs' feedback), along with eMBB input data \((\bm{s}(t))\), the number of URLLC packets arriving in each mini-slot $\tau$ in a TTI 
(i.e., $\bm{k}(t) = [k(t, 1), k(t, 2), \dots, k(t, 7)] $), and the actual ``selected" puncturing codes from the codebook for each mini-slot in TTI $t$ (i.e., $\bm{Y}(t) = [\bm{y}(t, 1), \bm{y}(t, 2), \dots, \bm{y}(t, 7)] $).
All this information is collected in the input buffer and used by the critics to update and refine policy network \(f_a\).

The purpose of using two critics, \(Q_1\) and \(Q_2\), is to independently evaluate the expected future rewards through their NN based on data in the input buffer. 
By choosing the minimum from these two critics' evaluations, overestimation bias is reduced, leading to greater stability and accuracy in the learning process.
We use two critics (instead of more) here to strike a balance between performance and complexity---adding more critics would increase computational overhead but cannot offer significant gains in learning efficiency.

\noindent \textbf{RT O-DU}\ \
In the O-DU, all possible numbers of URLLC packets that can arrive in a mini-slot $\tau$ (i.e.,  $1, 2, \cdots, \lfloor N/L \rfloor$)\footnote{The case when there is no URLLC packet arrival will not be considered since it does not involve URLLC puncturing.} 
will be considered.  
For each of the $\lfloor N/L \rfloor$ branches, the current policy network $f_a$ receives the scheduling vector $\bm{s}(t)$ along with $j$ (from 1 to $\lfloor N/L \rfloor$) URLLC packets.
The output of the $j$-th NN parameterizes a stochastic policy \( \pi_a^j \), which defines a set of probability density functions (PDFs) over the possible puncturing vectors (rather than deterministic one).
Specifically, the $j$-th NN produces a mean vector \(\bm{\mu_j}\) and a variance vector \(\bm{\sigma_j}\).
Both \( \bm{\mu_j} \) and \( \bm{\sigma_j} \) have a dimension of \( E \times 1 \), resulting in a total of $2E$ outputs.
Each element \( e \) in the \( \bm{\mu_j} \) and \( \bm{\sigma_j} \) vectors defines a Gaussian distribution, yielding a total of \( E \) Gaussian distributions.
Then, from the $e$-th Gaussian distribution, we take a sample for the number of punctured SCs for user \( e \).
These sampled values form a puncturing vector, denoted as $\bm{b}_j(t)$.

However, \(\bm{b}_j(t)\) may not be feasible, as the sampled values are unlikely to be  discrete and the number of preempted SCs for a given user \( e \) may exceed the total SCs allocated to that user $n_e(t)$. 
To address this issue, a policy wrapper called {\em Feasibility Enforcer\/} is employed, which adjusts \(\bm{b}_j(t)\) into a feasible solution, \(\bm{b}^{*}_j(t)\).

Then all $\lfloor N/L \rfloor$ puncturing vectors \(\bm{b}^{*}_j(t)\) are combined into a matrix 
\(\bm{B}^{*}(t)\), which is the {\em puncturing codebook\/} for the TTI.
The codebook has a dimension of  $E \times \lfloor \frac{N}{L} \rfloor$, where $E$ represents the number of eMBB users in the cell and 
$\lfloor \frac{N}{L} \rfloor$ is the total number of puncturing codes. 
This codebook is transmitted to the data plane (O-RU), which will be applied in each mini-slot, w.r.t. on {\em actual\/} the number of URLLC packet arrivals.

\subsubsection{Data Plane} 

The data plane of Cyrus+ operates in an  O-RU. 
The input to this plane includes the eMBB packets at TTI $t$ ($\bm{s}(t)$), the number of arriving URLLC packets in each mini-slot ($k(t, \tau)$), and the puncturing codebook ($\bm{B}^{*}(t)$) that comes from O-DU. 
In each mini-slot, a puncturing action is applied to the eMBB packet, based on the corresponding vector from the puncturing codebook.

\subsubsection{eMBB Feedback}

At each eMBB UE, the LDPC decoder is used to correct any corrupted bits (including those punctured by URLLC).
After LDPC, the Cyclic Redundancy Check (CRC) bits are checked to determine whether or not the eMBB packet transmission was successful. 
The results of the CRC check are then reported as feedback to the O-RU.

\section{Design Details} \label{sec:Design_Details}

In this section, we offer additional details to our design of Cyrus+. 

\subsection{Near-RT RIC}

\noindent \textbf{Reward Function:} \ \ \ As shown in Fig.~\ref{fig:ORANDiagram}, updates of policy $f_a$ happen periodically within the Near-RT RIC.
A reward function, denoted by \(r(t)\), is used to evaluate the effectiveness of \(\bm{Y}(t)\) on eMBB performance during TTI \(t\).
After error correction by the LDPC code, the performance of eMBB users is evaluated by checking the success of packet decoding at eMBB receivers, which is to verify the CRC bits in each packet at the end of the TTI.
Recall \(d_e(t)\) is a binary indicator representing the decoding outcome of eMBB user \(e\)’s packet at its receiver (see (\ref{eq:costmin_t})).
We define the reward at TTI $t$ as follows:  
\begin{equation}
    r(t) = \frac{1}{N}     
    \sum_{e \in \mathcal{E}} (d_e(t)-1) \cdot n_e(t)  \; . 
    \label{eq:rUF}
\end{equation} 
By definition, $r(t)$ represents the proportion of SCs (normalized by total $N$ SCs) that cannot be decoded successfully. 
The maximum attainable reward is $0$, achieved when all eMBB users successfully decode their received packets, while the minimum reward is $-1$, achieved when all eMBB users fail to decode their received packets. 
The primary objective of Cyrus+ is to achieve rewards as close to $0$ as possible.

The computed reward $r(t)$, along with 
\([\bm{s}(t), \bm{k}(t), \bm{Y}(t)]\) received from the data plane, are stored in the input buffer. 
Each entry in the input buffer contains the following data structure:  
\([\bm{s}(t), \bm{k}(t), \bm{Y}(t), r(t)]\).
When a new entry arrives and the input buffer is full, the oldest entry will be dropped (to leave room for the new entry).

\noindent \textbf{Critics Updates:} \ \ \ Denote $\phi_1$, $\phi_2$, and $\theta_a$ as the trainable NN parameters in the critics $Q_1$, $Q_2$, and policy network $f_a$ respectively.
Each critic is an NN with fully connected layers, independently estimates the expected discounted cumulative reward at the end of transmission, i.e., $E[\sum_{t=1}^M \lambda^{t-1}r(t)]$ , where $\lambda \in [0, 1]$ is the discount factor, $M$ represents the total number of mini-slots in each TTI ($7$).
Denote \(Q_{\text{targ}_i}\) as the {\em target $Q$-function} corresponding to critics \(i\), which is a separate NN to compute the target $Q$-values.
The target $Q$-function is a copy of the critic network but updates more slowly than the main critic.
These target networks help stabilize the learning of the critics, $Q_1$ and $Q_2$, and ensure the learning can converge eventually.  

Using data in the input buffer, the dual critics, $Q_1$ and $Q_2$, try to update their parameters $\phi_1$, $\phi_2$ to improve their prediction accuracy by minimizing the discrepancy between their predicted $Q$-values (i.e. their NN's output) and the ``target" $Q$-values (computed using separate NNs \(Q_{\text{targ}_i}\)).

For each training episode, 
a critic ($Q_1$ or $Q_2$) fetches $H$  samples randomly from the input buffer, 
\([\bm{s}(t_h), \bm{k}(t_h), \bm{Y}(t_h)]\), $h=1, 2, \cdots, H$.
For each of the $H$ samples, the training process for the critic involves a sequence of $M = 7$ steps corresponding to the seven mini-slots in the sample.
In each step $\tau$, 
$\tau =1, 2, \cdots, 7$,
the critic takes  $[\bm{s}(t_h), k(t_h, \tau), \bm{y}(t_h, \tau)]$ as its input.
In the final step ($\tau = M$) for each sample, the calculated reward from (\ref{eq:rUF}) is assigned to the critic.
During each training episode, each critic's NN is refined by performing the following gradient descent:
\begin{equation}
\begin{aligned}
\nabla_{\phi_i} \frac{1}{H \cdot M} \sum_{h=1}^{H} \sum_{\tau=1}^{M} 
\Big( Q_{\phi_i}\left([\bm{s}(t_h), k(t_h, \tau)], \bm{y}(t_h, \tau)\right) \\
- \hat{Q}\left(t_h, \tau) \right) \Big)^2 \quad \text{for}  \,\, i = 1, 2,
\end{aligned}
\label{eq:updatCritics}
\end{equation}
where \( Q_{\phi_i}(.) \) represents the Q-value computed by critic \( i \), and \( \hat{Q}(.) \) is the estimated Q-value derived from the target Q-function \( Q_{\text{targ}_i} \).
\(\hat{Q}(.)\) is computed using the reward function $r(t)$, the target Q-function \(Q_{\text{targ}_i}\), and entropy regularization, which we elaborate below.

As discussed in \ref{subsec:Cyrus_Overview}, the policy $\pi_a^j$ is a stochastic policy that provides a vector of Gaussian probability distributions (for $b_j(t)$) w.r.t a given state.
This stochasticity introduces variability in action selection (sampling randomly distributions). 
However, during training, the policy tends to rapidly converge to a narrowly concentrated distribution (suboptimal solution) by progressively reducing action variance and consistently selecting similar actions.
To mitigate this problem, entropy regularization is incorporated into both the policy and critics' learning processes.
Entropy regularization involves incorporating the policy's entropy into the learning process.
Entropy serves as a measure of uncertainty in the policy's decision-making and can be used to prevent  premature convergence to a suboptimal distribution. 
Entropy regularization, derived from $\pi_a^j$ (the probability distributions over actions), is formally defined as:
$$\mathbb{E}_{\bm{y}(t_h, \tau) \sim \pi_a^j}[ - \log \pi_{\theta_a}^j(\bm{y}(t_h, \tau) | [\bm{s}(t_h), k(t_h, \tau)])] $$
where $\pi_{\theta_a}^j$ represents the parameterized policy corresponding to the branch handling $k(t_h, \tau)$ URLLC packet arrivals ($j = k(t_h, \tau)$).
This entropy represents the expected (over the action distribution) negative logarithm of the policy’s probability density for the sampled action $\bm{y}(t_h, \tau)$ given the state $[\bm{s}(t_h), k(t_n, \tau )]$, based on the current policy $\pi_{\theta_a}^j$.
This quantifies the randomness in action selection for a given state and can be used to encourage exploration by penalizing low-entropy policies.
To regulate entropy, a coefficient \( \zeta \) is introduced in the entropy regularization term, which scales the contribution of entropy in the policy optimization objective.
A higher \( \zeta \) encourages greater exploration by increasing action randomness. 
Conversely, a lower \( \zeta \) reduces exploration, making the policy more deterministic as it prioritizes reward maximization over entropy.

This interplay between entropy and reward function is encapsulated in
\(\hat{Q}(.)\), which is computed as follows:
\begin{equation}
\begin{aligned}
\hat{Q}(t_h, \tau) &= 
\begin{cases}
\lambda \Bigg[\min\limits_{i=1,2} Q_{\text{targ}_i}\bigg([\bm{s}(t_h), k(t_h, \tau + 1)], \tilde{\bm{y}}(t_h, \\ 
\quad \tau+1)\bigg) - \zeta \log \pi_{\theta_a}^j\bigg(\tilde{\bm{y}}(t_h, \tau+1) \mid [\bm{s}(t_h), \\ 
\quad k(t_h, \tau + 1)]\bigg) 
\Bigg] \qquad \qquad  \qquad \tau < M -1 , \\
r(t) \qquad \qquad \qquad \qquad \qquad \qquad  \tau = M -1.
\end{cases}
\end{aligned}
\label{eq:target}
\end{equation}
In our problem setting, the reward $r(t)$ is only available at the end of the slot, meaning that intermediate steps do not directly receive rewards. 
To enable learning despite this delayed reward structure, we use the Q-target \( Q_{\text{targ}} \) for the expected cumulative reward at each intermediate step (\(\tau < M\)). 
At the final step (\(\tau = M\)), the actual reward \( r(t) \) is used, capturing the impact of all preceding actions on the final outcome.
The action $\tilde{\bm{y}}(t_h, \tau+1)$ in (\ref{eq:target}) represents the action sampled from the current policy (i.e., \( \tilde{\bm{y}}(t_h, \tau+1) \sim \pi_{\theta_a}^j(\cdot \mid [\bm{s}(t_h), k(t_h, \tau + 1)]) \)) instead of using data in the input buffer.
Since SAC is an off-policy algorithm that improves the policy while using experience from past interactions, sampling fresh actions ensures that the Q-target reflects what the current policy is expected to do in the next step, preventing the learning from being overly biased by outdated behavior.

\begin{algorithm}[t]
\caption{Cyrus+'s Near-RT RIC}\label{alg:NearRT_RIC}
\small
\begin{algorithmic}[1]
\State \textbf{Input:} $\bigl[\bm{s}(t),\, \bm{k}(t),\, \bm{Y}(t)\bigr]$ and eMBB users feedback
\State \textbf{Output:} Updated policy network $f_a$

    \State Use eMBB users feedback to calculate reward $r(t)$
    \State Create tuple $[\bm{s}(t), \bm{k}(t), \bm{Y}(t), r(t)]$
    \If{input buffer is not  full}
        \State Save $[\bm{s}(t), \bm{k}(t), \bm{Y}(t), r(t)]$
    \Else
        \State Discard oldest sample in input buffer 
        \State Save $[\bm{s}(t), \bm{k}(t), \bm{Y}(t), r(t)]$ in input buffer
        \State \textbf{Do}
            \State \hspace{1em} Fetch $H$ samples randomly from input buffer
            \State \hspace{1em} Update critics parameters $\phi_1$ and $\phi_2$ in $Q_1$ and $Q_2$ by using the gradient descent (\ref{eq:updatCritics})
            \State \hspace{1em} Update policy network $f_a$ using critics $Q_1$ and $Q_2$ by using gradient ascent (\ref{eq:updatePolicyPi})
            \State \hspace{1em}Send updated policy network $f_a$ to O-DU
            \State \textbf{EndDo}
    \EndIf

\end{algorithmic}
\end{algorithm}

Once the two critics $Q_1$ and $Q_2$ are updated following \ref{eq:updatCritics}, the policy network $f_a$ will also be updated, which we elaborate below.

\noindent \textbf{Policy Updates} \ \ \  
To optimize policy network \(f_a\), Cyrus+ utilizes the Q-functions estimated by the critics.
The policy \( f_a \) is updated with one step of gradient ascent
\begin{equation}
\begin{aligned}
\nabla_{\theta_a} \frac{1}{H \cdot M} \sum_{h=1}^{H} \sum_{\tau=1}^{M} \Big[ \min_{i=1,2} Q_{\phi_i}\bigg([\bm{s}(t_h), k(t_h, \tau)], \bm{\tilde{y}}_{\scriptscriptstyle \theta_a}(t_n, \tau)\bigg) \\
- \zeta \log \pi_{\theta_a}^j\bigg(\bm{\tilde{y}}_{\scriptscriptstyle \theta_a}((t_n, \tau) \mid [\bm{s}(t_h), k(t_h, \tau)]\bigg) \Big].
\end{aligned}
\label{eq:updatePolicyPi}
\end{equation}
Here, we compute this gradient of the action-selection objective w.r.t. 
$\theta_a$, the trainable parameters of the policy network $f_a$. 
By differentiating through the action $\bm{\tilde{y}}_{\scriptscriptstyle \theta_a}(t_n, \tau)$, we ensure that changes to $\theta_a$ directly influence the action generation process and improve the policy effectively.
Using (\ref{eq:updatCritics}) and (\ref{eq:updatePolicyPi}), the policy network $f_a$ and both critics, $Q_1$ and $Q_2$, are continuously updated in the Near-RT RIC and O-DU.

Algorithm~\ref{alg:NearRT_RIC} summarizes the main steps in  Near-RT RIC.

\subsection{RT O-DU}\label{subsec:RT_ODU}

\noindent \textbf{Feasibility Enforcer} \ \ \  
As discussed in Section~\ref{sec:Cyrus+} (under ``RT O-DU"), the policy network $f_a$ may produce infeasible puncturing strategy $\bm{b}_j(t)$, necessitating an additional mechanism to ensure feasibility.
Recall that infeasibility may come in two forms: 
(i) Required number of SCs for puncturing on an eMBB SCs exceeds the number of allocated SCs for that eMBB user.
(ii) The required number of SCs for preemption is not an integer.

We now address (i). 
To resolve this infeasibility, we aim to find a feasible puncturing vector (denote as \(\bm{\hat{b}}_{j}(t)\)) that is ``nearest" to the infeasible puncturing vector.
To achieve this, we employ the Kullback-Leibler (KL) divergence \cite{kullback1951:JSTOR:KL} to represent the distance (nearness) between \(\bm{\hat{b}}_{j}(t)\) and \(\bm{b}_{j}(t)\) and minimize this distance.
The KL divergence is given by
\begin{equation}
\label{eq:KL_def}
D_{\text{\tiny KL}}(\bm{b}_{j}(t) \parallel \bm{\hat{b}}_{j}(t)) = \sum_{e \in \mathcal{E}} m_{e}(t, .) \log \frac{m_{e}(t, .)}{\hat{m}_{e}(t, .)},
\end{equation}
where the notation ``$.$'' inside a parenthesis indicates any mini-slot 
and 
\( m_{e}(t, .) \) and \( \hat{m}_{e}(t, .) \) represent the infeasible and feasible number of SCs punctured on user \( e \), respectively. 
We now need to solve the following optimization problem: 
\begin{align}
\min \quad & D_{\text{\tiny KL}}(\bm{b}_{j}(t) \parallel \bm{\hat{b}}_{j}(t)) \label{eq:KL_OPT} \\
\text{s.t.} \quad & \hat{m}_{e}(t, .) \leq n_e(t) \quad \forall e \in \mathcal{E}, \notag \\
& \hat{m}_e(t, .) \geq 0 \quad \forall e \in \mathcal{E}, \notag \\
& \sum_{e \in \mathcal{E}} \hat{m}_e(t, .) = j \cdot L. \notag
\end{align}
The above problem is a convex optimization (with convex objective and linear constraints). 
Although standard convex optimizers can solve it, we employ the water-filling algorithm proposed in \cite{duchi2008:book:waterfilling, darroch1972:JSTOR:KLSolver} as it requires fewer iterations.

The solution to (\ref{eq:KL_OPT}), $\hat{m}_e(t, .)$'s, may still be infeasible as they are unlikely to be integers.
This is the second infeasibility issue (ii) that we discussed earlier. 
To address this issue, we need to round up/down each value $\hat{m}_e(t, .)$ while ensuring their total sum is $j \cdot L$.  
This can be done by using the Huntington-Hill method \cite{Balinski2010:intiger}, which systematically rounds numbers up/down in a way that maintains proportionality and ensures that the final sum remains exactly $j \cdot L$.
After this step, we obtain a 
feasible puncturing vector $\bm{b}_j^*(t)$.

As discussed in Section~\ref{sec:Cyrus+}, puncturing vectors $\bm{b}_j^*(t)$ are combined to form the codebook $\bm{B}^*(t)$, which is applied in the O-RU.
Algorithms \ref{alg:RT_ODU} and \ref{alg:RT_ORU} summarize the steps used in Cyrus+ for RT O-DU and RT O-RU, respectively.

\begin{algorithm}[t]
\caption{Cyrus+'s RT O-DU}\label{alg:RT_ODU}
\small
\begin{algorithmic}[1]
\State \textbf{Input:} $f_a$ (initialized with $f_0$) and eMBB scheduling $\bm{s}(t)$
\State \textbf{Output:} Codebook $\bm{B}^*(t)$ 
\For{TTI $t = 1, 2, 3, \dots$}
    \State \textbf{Do} in parallel for $j=1,\dots,\lfloor N/L \rfloor$
        \State \hspace{1em} Sample $\bm{b}_j(t) \sim \pi_{\tiny \theta_a}^j(.\mid[\bm{s}(t), j])$ 
        \State \hspace{1em} Using Feasibility Enforcer to find $\bm{b}_j^*(t)$
    \State \textbf{EndDo}
    \State Send codebook $\bm{B}^*(t)$ to RT O-RU
\EndFor
\end{algorithmic}
\end{algorithm}

\begin{algorithm}[t]
\caption{Cyrus+'s RT O-RU}\label{alg:RT_ORU}
\small
\begin{algorithmic}[1]
\State \textbf{Input:} Codebook $\bm{B}^*(t)$
\State \textbf{Output:} $\bigl[\bm{s}(t),\, \bm{k}(t),\, \bm{Y}(t)\bigr]$ and eMBB feedback $d_e(t)$ for $e\in \mathcal{E}$

\State For $[\bm{s}(t),\, \bm{k}(t)]$, construct $\bm{Y}(t)$ based on $\bm{B}^*(t)$

    \State Send tuple $\bigl[\bm{s}(t),\, \bm{k}(t),\, \bm{Y}(t)\bigr]$ and $d_e(t)$'s to Near-RT RIC
\end{algorithmic}
\end{algorithm}

\subsection{Non-RT RIC}\ \
\label{subsec:setting_pi_0_in_non_RT_ric}

While the pre-training process in the Non-RT RIC enjoys much relaxed time scale, the final trained model, $f_0$, however, will shares the same NN architecture as $f_a$ in the O-DU.
So the first question is: 
What does the NN architecture for $f_0$ (or $f_a$) should be?  

One of the most important factors in designing $f_a$ is meeting its timing constraint in the O-DU.
Note that all computations required to determine the new puncturing codebook \(\bm{B}^{*}(t)\), including the forward pass through policy \(f_a\) and feasibility enforcer computation, must be completed within one TTI (as short as $125 \mu s$).
To meet this stringent timing requirement, we must avoid using too many hidden layers and neurons for policy $f_a$.
In O-DU, we find that the forward pass must be completed within $80$ $\mu s$ as the feasibility enforcer takes $10 - 30 \mu s$.
Based on a trial-and-error approach, we find that 
a three-layer structure, with 
$128$ neurons for the hidden layer would meet the required forward pass time requirement.

However, this simple NN structure for $f_0$ (same as $f_a$) also introduce a new problem. 
It poses a severe limit on the agent’s ability to fully learn environmental features.
Our experiments (see Section~\ref{sec:Performance}) revealed that even after $2 \times 10^5$ iterations, there was no sign of convergence in the reward function.
This difficulty is mainly because the simple three-layer structure of $f_0$ 
cannot cope with the rapidly changing eMBB scheduling $\bm{s}(t)$ used in each TTI and varying number of URLLC packet arrivals in each mini-slot during the training process.

To address this convergence issue, 
we propose to employ advanced curriculum learning (ACL) \cite{Yoshua:Curriculmlearning:ICML, Portelas20:AutomaticCL:survay, Carlos17:ReverseCurriculum}.
The idea of ACL is to start with simple tasks in training (e.g., datasets with low complexity).
Once the policy network $f_0$ becomes proficient in solving these simple tasks, ACL then steps up with more challenging tasks.  
Specifically, Cyrus+ takes the following steps in ACL: 
\begin{itemize} 

\item 
We start with the simplest scenario with one URLLC arrival in a TTI where the URLLC packet may fall into any one of the seven mini-slots.  
We also generate a specific eMBB SC allocation $\bm{s}(t)$.
For this specific input setting of eMBB and URLLC $[\bm{s}(t), \bm{k}(t)]$, instead of using it for just one TTI, we repeat it for many times over a window (say 50 TTIs).  
During this window, different puncturing vector $\bm{b}_j^*(t)$ generated by the policy network $f_0$ is applied in each TTI.  
For the next window, we repeat the same exercise, with a different eMBB SC allocation and still one URLLC packet (which may fall into a different mini-slot). 
We will repeat these windows until the reward function $r(t)$ converges.

\item Once the reward function $r(t)$ converges for the above simplest case (a window with one URLLC arrival), we step up the training complexity with a smaller window size and more URLLC packets. 
Specifically, we increase URLLC packet arrivals in a TTI to two (randomly falling into any mini-slots) and decrease the window size, say to 40 TTIs. 
Again, we repeat this same input for as many windows as needed until the reward function $r(t)$ converges.

\item  The process continues until the window size degenerates into just one TTI, where we have one new eMBB SC allocation for each TTI and the maximum number of URLLC packets that can be allowed to enter without causing the average reward to drop to $-1$ (i.e., on average none of the eMBB receivers at the UE side can decode any packet successfully).\footnote{Such a maximum number of URLLC packets can be found empirically through experiments, as we shall show in our experimental results (Section~\ref{sec:Performance}).}


\end{itemize}

\section{Implementation} \label{sec:Implementation}

\begin{figure}[t]
    \centering
    \includegraphics[width=\linewidth]{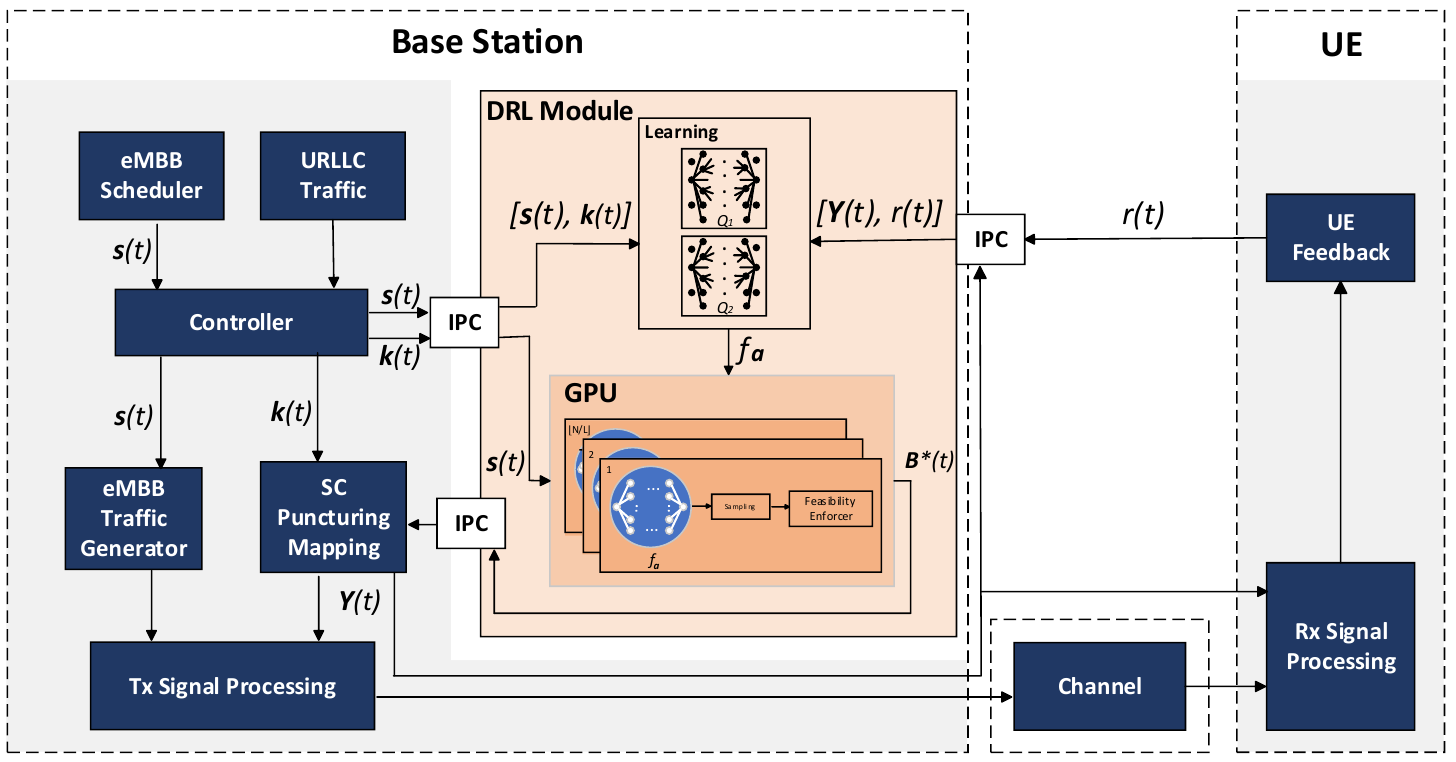}
    \caption{Architecture diagram of the NR simulator.}
    \label{fig:Implementation1}
\end{figure}

In this section, we present our implementation of Cyrus+ as a simulator. 
Our hardware platform is a DGX V100 with a 20-core Intel Xeon CPU E5-2698 v4.
We use two NVIDIA Tesla V100 GPUs in the DGX station.
On the software side, 
we employ MATLAB R2023b with 5G NR Toolbox \cite{MathWorks2023:online:5GToolbox}
to implement physical and MAC layer and TensorFlow version 2.15.0 \cite{abadi2016:arxiv:tensorflow} with tf-agent version 0.19.0 to implement our DRL agent.  
Figure~\ref{fig:Implementation1} illustrates the architectural dragram of our NR simulator, which consists of three components: BS, channel, and UE.
we describe each component below. 

\subsection{Base Station}\label{subsec:Simulator}
The BS includes the MAC-PHY layer, which manages resource allocation and transmission, and the DRL module, which controls the puncturing process.
We first provide a detailed explanation of the MAC-PHY layer in this section, while the DRL module is discussed in Section~\ref{subsec:DRL}.

The MAC-PHY layer of the BS consists of several modules for downlink processing, each of which is described below.

\noindent \textbf{eMBB Scheduling} \ \ \ 
This module schedules eMBB users and allocates the available SCs to each user.
This scheduling is performed using the Proportional Fair (PF) scheduler \cite{Kelly98:JonOprational:PFScheduler}.
After scheduling, an MCS is selected for each eMBB user.
The 5G Toolbox is used to generate both the scheduling vector and the MCS assignment.

\noindent \textbf{URLLC Traffic}\ \ \  
This module generates traffic profile information about incoming URLLC packets for the next TTI.
This module is written in MATLAB.
 
\noindent \textbf{Controller} \ \ \ 
This module manages the number of URLLC packet arrivals within a given mini-slot, which is limited by the available SCs.
It buffers excess URLLC packets (those exceeding the bandwidth limit) 
for transmission in subsequent mini-slots.
From this module, the eMBB scheduler $\bm{s}(t)$ and the modified URLLC demand $\bm{k}(t)$ are sent to the DRL module via Inter-Process Communication (IPC).
 
\noindent \textbf{IPC}\ \ \
This module facilitates communication between the MAC-PHY layer and the DRL module. 
It transfers $\bm{s}(t)$, $\bm{k}(t)$, $\bm{Y}(t)$, and $r(t)$ from the MAC-PHY layer to the DRL module and sends the codebook $\bm{B}^*(t)$ back to the MAC-PHY layer.

\noindent \textbf{eMBB Traffic Generator} \ \ \ 
In this module, we use  $\bm{s}(t)$ and MCS level for each user from the controller to generate actual data bits for eMBB users.
The bit sequence is generated randomly, and CRC bits are appended to the eMBB data bits.

\noindent \textbf{SC Puncturing Mapping}\ \ \ 
This module runs in parallel with the eMBB traffic generator.
In this module, we receive the codebook $\bm{B}^*(t)$ from the DRL module.
Using the computed puncturing codebook $\bm{B}^*(t)$ and the number of arriving URLLC packets profile $\bm{k}(t)$, we determine the puncturing matrix $\bm{Y}(t)$.
Since this paper focuses on the performance of eMBB users, we can use any bit sequence for URLLC users
(which will be all zeros in our simulator). 
The matrix $\bm{Y}(t)$ is then sent to the Tx signal processing module to puncture URLLC users into eMBB users. 
Additionally, $\bm{Y}(t)$ is forwarded to the DRL module and the Rx receiver processing module to inform each UE about the punctured portions of its data.

\noindent \textbf{Tx Signal Processing} \ \ \ 
In this module, we implement a downlink Tx signal processing chain at the BS.
The transmission process includes LDPC encoding, rate matching, resource puncturing, scrambling, modulation mapping, and OFDM modulation.
We utilize standard functions from the 5G Toolbox for this process, except for resource puncturing, which is implemented by us.
For resource puncturing, 
eMBB data is modified based on the selected puncturing matrix $\bm{Y}(t)$.
The final data is then transmitted to the channel toward the UEs.

\subsection{DRL Module}
\label{subsec:DRL}

In Fig.~\ref{fig:Implementation1}, the DRL module consists of two sub-modules.
In the upper sub-module, the input buffer (not shown in the figure) receives scheduling vector \( \bm{s}(t) \) and URLLC demand \( \bm{k}(t) \) from the BS.
This input buffer is a fixed-size circular buffer that stores incoming data sequentially.
Upon receiving \( \bm{s}(t) \) and \( \bm{k}(t) \), the buffer writes these values at the current index position and then awaits the corresponding puncturing vector \( \bm{Y}(t) \) and the reward \( r(t) \) from the UEs.
Once all necessary information is available, the buffer entry is completed and the index pointer is incremented to the next position.
If the buffer reaches its maximum capacity, the oldest data is overwritten according to a First-In-First-Out (FIFO) policy, using circular indexing.
Next, both critics' NN parameters, $\phi_1$ and $\phi_2$, are updated using samples from the input buffer.
Each critic is implemented as a fully connected NN with two hidden layers.
Finally, the critics update the policy network $f_a$ and send it to the lower GPU sub-module.

In the GPU sub-module, 
based on the received scheduling vector \(\bm{s}(t)\) (from MAC-PHY layer), 
we generate a puncturing vector \(\bm{b}_j(t)\) for 
each possible URLLC packet arrivals (from $j=1$ to $\lfloor \frac{N}{L} \rfloor$.
Then, we apply a feasibility enforcer to each \(\bm{b}_j(t)\) to compute a feasible puncturing vector \(\bm{b}_j^*(t)\).
After this step, we put the vectors $\bm{b}_j^*(t)$, $j = 1, 2, \cdots, \lfloor \frac{N}{L} \rfloor$ together as the puncturing codebook \(\bm{B}^{*}(t)\) and transmit this codebook to the MAC-PHY layer via IPC.
We design and implement this sub-module on a GPU which we elaborate below.

\begin{figure}[t]
    \centering
    \includegraphics[width=\linewidth]{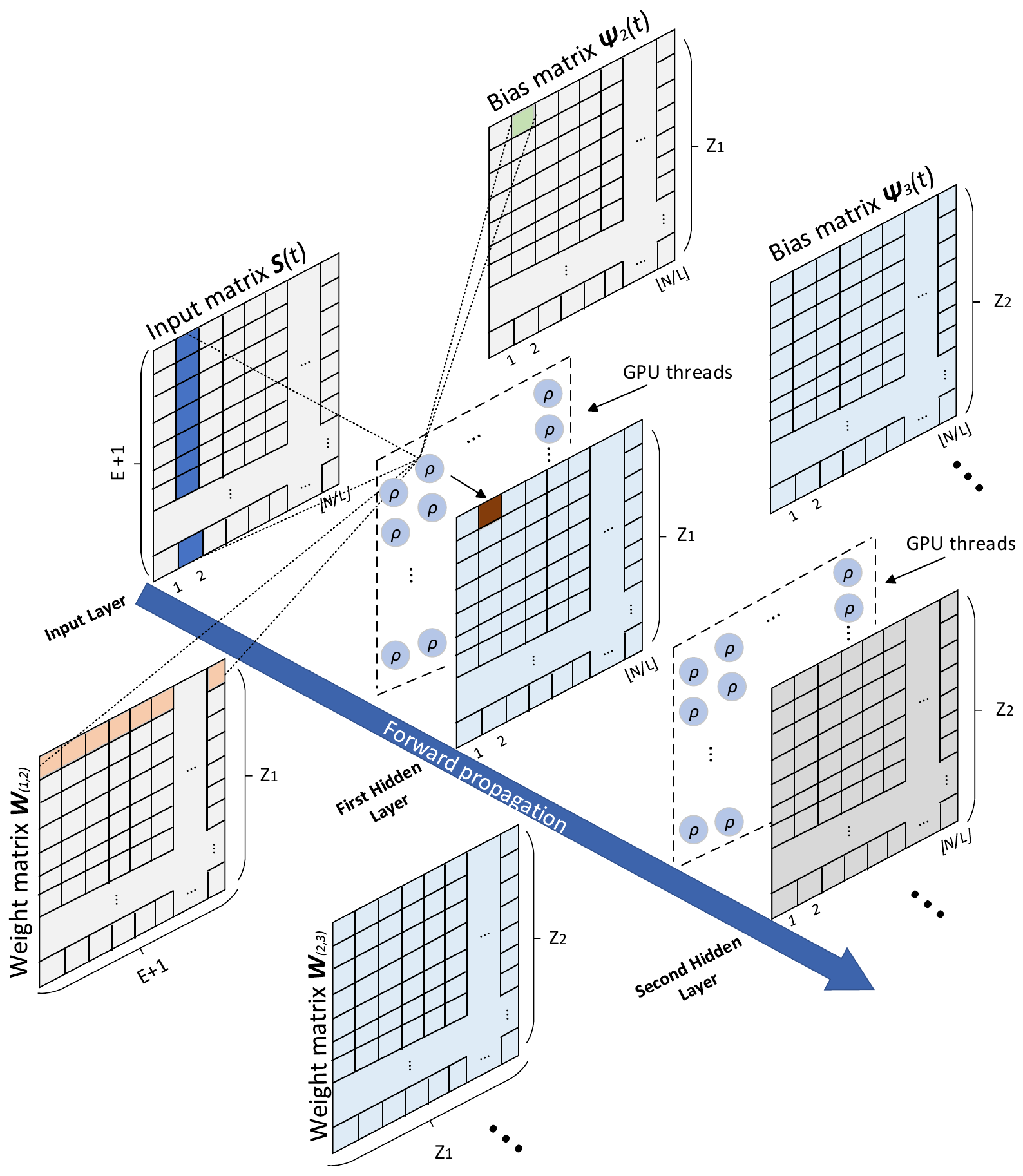}   
    \caption{Concurrent forward propagation for all possible URLLC packet arrivals ($\lfloor \frac{N}{L} \rfloor$).}    
    \label{fig:GPUImplementation}
\end{figure}

\noindent \textbf{GPU Implementation} \ \ \
Recall that the input to the policy network $f_a$ in each $j$-th branch is $[\bm{s}(t), j]$ (see Fig.~\ref{fig:ORANDiagram}).
The forward propagation for all $j \in \{1, \cdots, \lfloor \frac{N}{L} \rfloor\}$ possible URLLC packet arrivals can be executed in parallel using matrix operations on a GPU \cite{NVIDIA2024:online:cuDNN}.
Figure~\ref{fig:GPUImplementation} illustrates this concurrent forward propagation.
Specifically, we combine all $[\bm{s}(t), j]$ inputs into a matrix
$\bm{S}(t) = \bigg[[\bm{s}(t), 1]; [\bm{s}(t), 2]; \cdots; [\bm{s}(t), \lfloor \frac{N}{L} \rfloor] \bigg]^T$, where ``$T$" represents the transpose operation.
The matrix $\bm{S}(t)$ has \(\lfloor \frac{N}{L} \rfloor\) columns, with each row of dimension $E+1$.
In $f_a$, denote $W_{(1, 2)}$ as the weight matrix connecting the input layer to the first hidden layer, which has a dimension $Z_1 \times (E+1)$, where $Z_1$ is the number of neurons in the first hidden layer.
Denote \(\bm{\psi}_{(2)}\) as the bias associated with the second layer, having dimension $Z_1 \times \lfloor \frac{N}{L} \rfloor$, and \(\rho(.)\) as the activation function used for the hidden layers. 

The forward propagation from the input layer to the first hidden layer is then computed as
\(\rho(\bm{W}_{(1, 2)} \cdot \bm{S}(t) + \bm{\Psi}_{(2)})\), where \(\bm{\Psi}_{(2)} = [\bm{\psi}_{(2)}; \bm{\psi}_{(2)}; \cdots; \bm{\psi}_{(2)}]^T\) is a matrix with bias vectors. 
As illustrated in Fig.~\ref{fig:GPUImplementation}, we employ a two-dimensional array of GPU threads to perform this computation.
Once all threads complete the computation of
\(\rho(\bm{W}_{(1, 2)} \cdot \bm{S}(t) + \bm{\Psi}_{(2)})\), the forward propagation proceeds to the next layer.

The above process continues from one layer to the next, 
until the final output is obtained. 
In the final output matrix, each column represents the parameters (means and variances) of the probability distributions to obtain the puncturing vector $\bm{b}_j(t)$.
Specifically, 
each column consists of $2E$ elements $[\mu_1, \cdots, \mu_{\tiny E}, \sigma_1, \cdots, \sigma_{\scriptsize E}$].
The $j$-th element of $\bm{b}_j(t)$ is obtained by sampling a Gaussian distribution using  $[\mu_j, \sigma_j]$.
For sampling, we employ the reparameterization trick \cite{kingma13:arXiv:ParametrizeTrick}. 
The resulting output matrix, where each column corresponds to  $\bm{b}_j(t)$, $j = 1, \cdots, \lfloor \frac{N}{L} \rfloor$, is then processed in parallel through the feasibility enforcer.
The feasibility enforcer is specifically designed to enforce output feasibility by addressing two distinct types of infeasibility (see Section~\ref{subsec:RT_ODU}). 
It is implemented as a class with two dedicated methods, one for puncturing SCs more than the allocated SCs to each user and the other for making the number of punctured SCs an integer.
The resulting feasible puncturing vectors $\bm{b}_j^*(t)$'s form the final output matrix representing  the codebook $\bm{B}^*(t)$.

\subsection{Channel} \label{Channel}
We use 5G Toolbox to generate a channel that simulates frequency and time-domain fading along with noise. 
We use the clustered delay line (CDL) channel to model multipath channel fading.

\subsection{UE} \label{subsec:UES}
A UE's receiver consists of two modules, which we describe as follows.

\noindent \textbf{Rx Signal Processing} \ \ \  
This module implements a receive signal processing chain at a UE.
It processes eMBB data received from the channel.
In this module, along with the eMBB data received from the channel, puncturing information $\bm{Y}(t)$ is also received, enabling the system to identify punctured segments during the decoding process.
The receiver-side processing involves multiple stages, including synchronization, OFDM demodulation, soft demodulation, descrambling, puncturing flushing, and LDPC decoding. 
For this process, we utilize standard functions from the 5G Toolbox, except for puncturing flushing. 
The purpose of puncturing flushing is to use the positions of punctured bits ($\bm{Y}(t)$) on eMBB packet to inform the LDPC decoder.
Since the eMBB bits at these positions are now unknown after URLLC puncturing, we set 
the confidence level of these bits to NULL after the soft demodulator, indicating we have no idea what these bits could be. 
With this information as input, 
the LDPC decoder will remove these missing bits during decoding 
\cite{ha2003:Book:LDPCPunctruing}.

\noindent \textbf{UE Feedback} \ \ \ 
In this module, a bitwise comparison is performed between the CRC bits extracted after decoding and the CRC bits originally attached at the BS.
Based on the comparison, we set $d_e(t)$.
The reward function $r(t)$ is then computed and transmitted to the DRL module along with the puncturing matrix $\bm{Y}(t)$ via IPC.

\subsection{Non-RT RIC} \label{subsec:NonRT_Implementation}

We now describe the training process in Non-RT RIC (not shown in Fig.~\ref{fig:Implementation1}) to determine the initial parameters of the policy network $f_0$.
The training structure is the same as the above implementations for the BS, channel, and UE, 
except that we need to modify the behaviors of eMBB scheduler and URLLC traffic modules,
which we describe as follows.

As explained in Section~\ref{subsec:setting_pi_0_in_non_RT_ric}, we use the ACL method during the training of the policy network $f_0$.
To implement ACL, 
we randomly generate both eMBB scheduling $\bm{s}(t)$ and MCS level for a TTI and repeat the same eMBB scheduling $\bm{s}(t)$ and MCS for the entire window (50 TTIs).  
Similarly, in the URLLC traffic module, URLLC packet arrival is also repeated over the entire training window after its generation.
We start with a single URLLC packet arrival and gradually increases the number of URLLC  packets over time.

\section{Performance Evaluation} \label{sec:Performance}

In this section, we evaluate the performance of Cyrus+ based on the implementation in Section~\ref{sec:Implementation}.
We organize this section as follows. 
In Section~\ref{subsec:benchmark}, we discuss the benchmarks that we use in 
evaluating Cyrus+.
In Section~\ref{subsec:case_study}, we present a case study. 
In Section~\ref{subsec:varying_parameters}, we study Cyrus+'s behavior under varying system parameters.  

\begin{table}[]
    \centering
    \caption{Parameter settings in the simulator.}
    \renewcommand{\arraystretch}{1}  
    \label{tab:Parameters}
    \small
    \begin{tabular}{|l|l|}
        \hline
        \rowcolor{gray!15} \multicolumn{2}{|c|}{\textbf{BS}} \\ \hline
       System bandwidth   & 50 MHz \\ \hline
          Carrier frequency  & 3750 MHz \\ \hline
        
        NR numerology      & 2  \\ \hline
        SC spacing      & 60 kHz \\ \hline
        
        Total SCs ($N$)  & 780 \\ \hline

        Tx power        & 46 dBm \\ \hline
        Antenna config.        & 16 Tx antenna \\ \hline

        \rowcolor{gray!15} \multicolumn{2}{|c|}{\textbf{Channel}} \\ \hline
        Pathloss model     & 3D UMa NLOS \\ \hline
        Fading channel     & CDL-C with 300 ns RMS delay spread \\ \hline

        \rowcolor{gray!15} \multicolumn{2}{|c|}{\textbf{UE}} \\ \hline
        eMBB antenna        & 2 Rx antenna \\ \hline
        URLLC antenna       &  1 Rx antenna \\ \hline
        Noise floor        & -84 dBm \\ \hline
        Channel estimation & Ideal \\ \hline

        \rowcolor{gray!15} \multicolumn{2}{|c|}{\textbf{DRL Module}} \\ \hline
        Discount factor ($\lambda$)  & 0.95 \\ \hline
        Learning rate     & $3 \times 10^{-4}$ \\ \hline
        Input buffer size & $2 \times 10^4$ \\ \hline
        Sampling size ($H$)      & $256$ \\ \hline
        Target update rate        & $0.005$ \\ \hline
    \end{tabular}
\end{table}

\begin{table}
\centering
\caption{Locations of eMBB and URLLC UEs.}
\label{table:locations}
\renewcommand{\arraystretch}{1}
\small
\begin{tabular}{|l|llll|}
\hline
UE Type & \multicolumn{4}{l|}{Coordinates (in (m, m))}  \\ \hline
\multirow{5}{*}{eMBB}  
& \multicolumn{1}{l|}{1} & \multicolumn{1}{l|}{(-164.2, 841.0)} 
& \multicolumn{1}{l|}{6}  & (-754.9, 509.7)  \\ \cline{2-5} 
& \multicolumn{1}{l|}{2} & \multicolumn{1}{l|}{(1.2, 581.4)}   
& \multicolumn{1}{l|}{7}  & (302.0, 445.6)   \\ \cline{2-5} 
& \multicolumn{1}{l|}{3} & \multicolumn{1}{l|}{(-813.3, 98.1)}  
& \multicolumn{1}{l|}{8}  & (430.8, 96.3)    \\ \cline{2-5} 
& \multicolumn{1}{l|}{4} & \multicolumn{1}{l|}{(958.1, 159.3)}  
& \multicolumn{1}{l|}{9}  & (67.6, -155.8)   \\ \cline{2-5} 
& \multicolumn{1}{l|}{5} & \multicolumn{1}{l|}{(573.9, -412.3)} 
& \multicolumn{1}{l|}{10} & (308.7, -530.3)  \\ \hline
\multirow{6}{*}{URLLC} & \multicolumn{1}{l|}{1} & \multicolumn{1}{l|}{(-15.6, 688.6)}  
& \multicolumn{1}{l|}{7}  & (-308.0, 576.6)  \\ \cline{2-5} 
& \multicolumn{1}{l|}{2} & \multicolumn{1}{l|}{(-882.5, 398.3)}
& \multicolumn{1}{l|}{8}  & (-546.7, 235.9)  \\ \cline{2-5} 
& \multicolumn{1}{l|}{3} & \multicolumn{1}{l|}{(299.9, 260.2)} 
& \multicolumn{1}{l|}{9}  & (283.1, 166.3)   \\ \cline{2-5} 
& \multicolumn{1}{l|}{4} & \multicolumn{1}{l|}{(-741.3, -31.6)}
& \multicolumn{1}{l|}{10} & (-712.9, 156.6)  \\ \cline{2-5} 
& \multicolumn{1}{l|}{5} & \multicolumn{1}{l|}{(756.0, -169.7)}
& \multicolumn{1}{l|}{11} & (8.6, -390.2)    \\ \cline{2-5} 
& \multicolumn{1}{l|}{6} & \multicolumn{1}{l|}{(255.9, -426.6)} 
& \multicolumn{1}{l|}{12} & (-197.2, -927.9) \\ \hline
\end{tabular}
\end{table}

\subsection{Benchmarks}
\label{subsec:benchmark}

For performance comparison, we consider the following two puncturing techniques that were introduced in Section~\ref{sec:Introduction}.
\begin{itemize}

\item {\bf Resource Proportion (RP)\/} punctures the required number of URLLC SCs on the eMBBs in proportional to the size of eMBB packets 
\cite{Anand18:INFOCOM:Joint_URLLC_eMBB}.
The idea is to spread out the potential adverse impact (due to URLLC puncturing) fairly across the eMBB packets.

\item {\bf Smallest eMBB First (SeF)\/} selects the smallest eMBB packet (in terms of its SCs) and puncture all its SCs first, and then move on to the second smallest eMBB packets, until all URLLC SCs are taken care of.  
The idea is to sacrifice the smallest eMBB packsts first by offering priority protection to the largest eMBB packets, in the hope that the adverse impact on the overall system throughput can be minimized.  
\end{itemize}

\subsection{A Case Study}\label{subsec:case_study}

\subsubsection{Experimental Setting} 

\begin{figure}[t]
    \centering
    \includegraphics[width=0.8\linewidth]{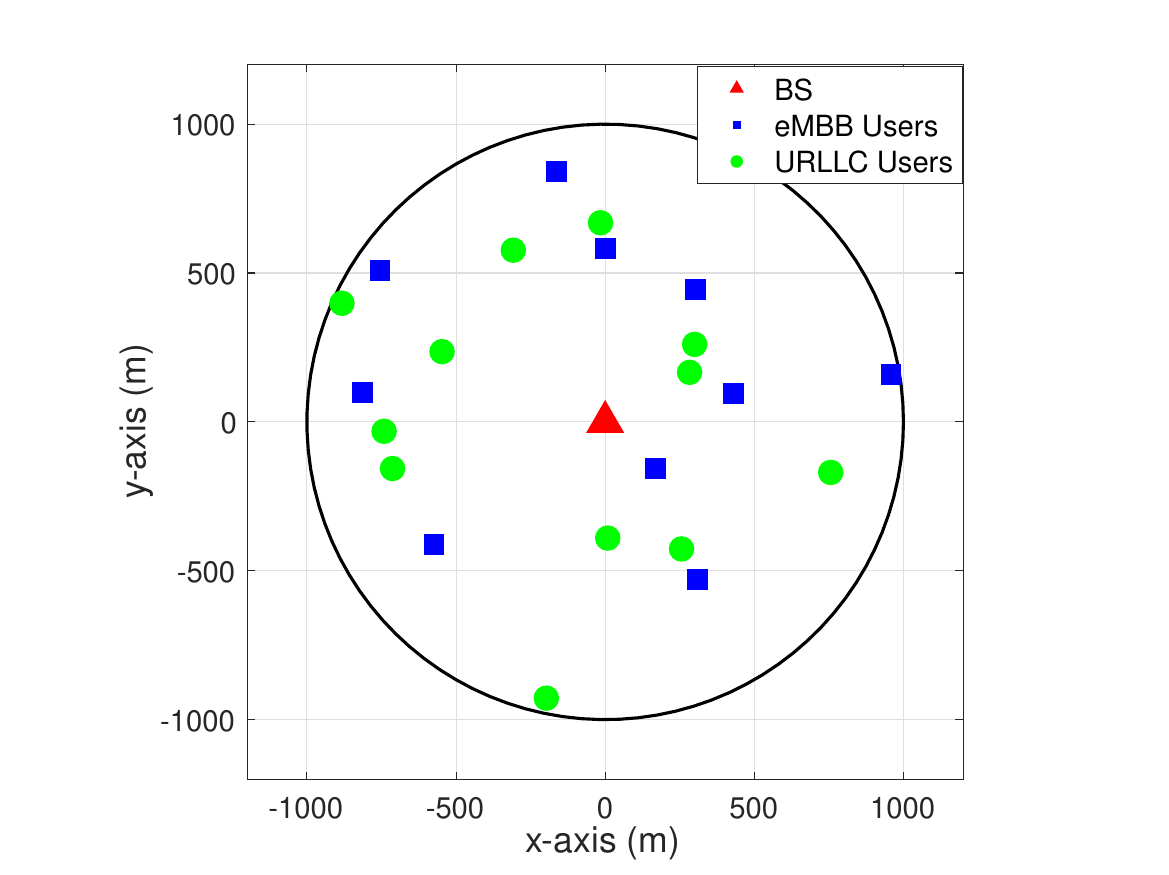}
    \caption{ A network topology with $10$ eMBB users and $12$ URLLC users.}
    \label{fig:NetworkTopology}
\end{figure}

Figure~\ref{fig:NetworkTopology} shows the network topology in our case study.  
We consider a 5G small cell with a radius of 1 km.  
The 5G NR BS is located at the center of the small cell and there are $10$ eMBB and $12$ URLLC UEs randomly deployed within the cell. 
The locations of the eMBB and URLLC UEs are given in Table~\ref{table:locations}.

The key parameters for each component of the simulator are given in Table~\ref{tab:Parameters}.
At the BS, data transmission employs the Zero Forcing (ZF) precoding technique.
BS uses a single codeword for both eMBB and URLLC transmissions.
For eMBB users, the BS employs the PF scheduling to select eMBB UEs for resource allocation in each TTI, where the resource granularity is defined at the  
resource block (RB) level, or equivalently 12 SCs.
We assume a full-buffer traffic model, where each eMBB user always has data to transmit. 
For each URLLC UE, the packet arrival process in each mini-slot follows a Bernoulli process with a probability $p = 0.08$.
For each URLLC packet, it has a fixed size of $50$ bytes and employs QPSK modulation with a code rate $1/3$ \cite{3gpp:tr38824:urllc}.  
Under this setting, each URLLC packet is encoded into $1200$ coded bits ($1/3$ coding rate), which are then mapped to $600$ QPSK symbols.
These symbols are allocated across $300$ SCs (\(L = 300\)) per mini-slot (as each mini-slot contains $2$ OFDM symbols).

In the DRL module, we employ a NN with one densely connected hidden layer with $128$ neurons for the policy network \(f_0\) (similar with $f_a$).
We use the Rectified Linear Unit (ReLU) \cite{Nair10:ICML:ReLU} for the activation function in this layer.
To ensure that the sampled actions remain within the range $[0, 1]$, we apply a hyperbolic tangent (\(\tanh\)) function at the final layer after sampling.
To optimize the NN’s weights, we employ the Adam optimizer \cite{kingma2017:Arxiv:adam}, configured with a learning rate of \(3 \times 10^{-4}\).
The critics' architecture (\(Q_1\) and \(Q_2\)) consists of two dense hidden layers, each containing $256$ neurons, employing the ReLU activation function.
Both critics are similarly trained using the Adam optimizer with a learning rate of \(3\times10^{-4}\).
We apply soft updates to the target networks with a coefficient of $0.005$.
The weights for both the policy network $f_0$ (and $f_a$) and the critics, $Q_1$ and $Q_2$, are randomly initialized.
For the input buffer, we set the buffer size to $2 \times 10^4$ samples.
The critics $Q_1$ and $Q_2$ fetch batches of $H = 256$ samples randomly from the input buffer during training. 

\subsubsection{Results} 
We now present detailed results from the case study. 

\noindent \textbf{Goodput Performance}\ \ \ 
To compare the goodput among  
Cyrus+, RP, and SeF, we run each 
algorithm over $5,000$ TTIs. 
We find that Cyru+ achieves the highest goodput at $49.852$ Mb/s, while RP and SeF achieves  $16.430$ Mb/s and $27.384$ Mb/s, respectively.

\begin{figure}[t]
    \centering
    \includegraphics[width=\linewidth]{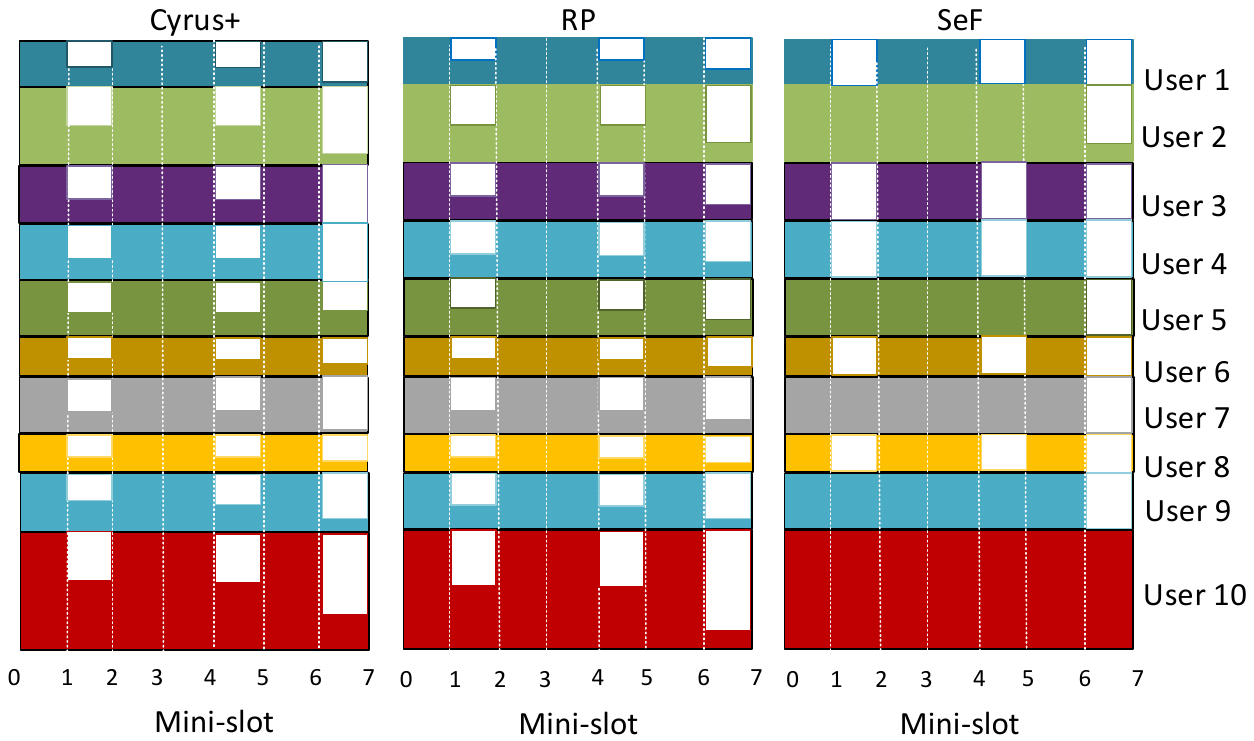}
    \caption{URLLC puncturing behviaor at TTI 300 under Cyrus+, RP, and SeF algorithms.  Punctured SCs are shown in white.}
    \label{fig:puncturing}
\end{figure}

To see how URLLC puncturing is done under different algorithms, we zoom into one specific TTI, say 300.  
The scheduling vector for the 10 eMBB users in this TTI is 
$\bm{s}(t) = [60, 108, 72, 72, 72, 48, 72, 48, 96, 132]$, all in SCs. 
In this TTI, there is one URLLC packet arrival with 300 SCs in the second mini-slot, 
another URLLC packet with 300 SCs in the fifth mini-slot, 
and two URLLC packets with 600 SCs total in the seventh mini-slot.
Figure~\ref{fig:puncturing} shows the eMBB packet for the TTI and the URLLC puncturing (in white) in each mini-slot under the three algorithms.  
Table~\ref{tab:scheduling_puncturing} lists the 
number of punctured SC in mini-slots 2, 5, and 7 under the three algorithms.
For example, for eMBB 1, under Cyrus+, the number of SCs punctured in mini-slots 2, 5, and 7 is represented as $(28, 28, 48)$. 
The last row in the table shows the sum of SCs for all the 10 rows above it.

The difference in puncturing strategies significantly impact the ability of eMBB users to decode data correct. 
For this specific TTI, 
eMBB packets 1, 2, 5, 6, 8, 9, and 10 can be decoded successfully under Cyrus+ 
(shown in shaded boxes in the corresponding column in Table~\ref{tab:scheduling_puncturing}), 
yielding a goodput of 59.339 kb in this TTI. 
In contrast, under RP, only eMBB packets 1, 2, 6, and 8 can be decoded successfully, 
yielding a goodput of 
23.092 kb; while under 
SeF, only eMBB packets 2, 5, 7, 9, and 10 can be decoded successfully, 
yielding a goodput of 51.897 kb.

\begin{table}
    \centering
    \caption{URLLC puncturing behavior at TTI 300 under Cyrus+, RP, and SeF.
    eMBB packets that can be decoded successfully are marked by a box.}  
    \label{tab:scheduling_puncturing}
    \scriptsize 
    \begin{tabular}{|l|l|l|l|l|}
    \hline
    eMBB  & \begin{tabular}[c]{@{}l@{}}$\bm{s}(t)$\\ (in SC)\end{tabular} & \begin{tabular}[c]{@{}l@{}}Cyrus+\\ (in SC)\end{tabular} & \begin{tabular}[c]{@{}l@{}}RP\\ (in SC)\end{tabular} & \begin{tabular}[c]{@{}l@{}}SeF\\ (in SC)\end{tabular}\\ \hline
    1 & 60 & \cellcolor[HTML]{E6FFE6}(28, 28, 48) & \cellcolor[HTML]{E6FFE6}(23, 23, 47) & (60, 60, 60)  \\ \hline
    2 & 108  & \cellcolor[HTML]{E6FFE6}(41, 41, 92) & \cellcolor[HTML]{E6FFE6}(41, 41, 83) & \cellcolor[HTML]{E6FFE6}(0, 0, 60) \\ \hline
    3   & 72   & (33, 33, 71)  & (28, 28, 55) & (72, 72, 72) \\ \hline
    4 & 72   & (27, 27, 71)  & (28, 28, 55) & (72, 72, 72) \\ \hline
    5 & 72 & \cellcolor[HTML]{E6FFE6}(28, 28, 31) & (28, 28, 55) & \cellcolor[HTML]{E6FFE6}(0, 0, 72) \\ \hline
    6   & 48 & \cellcolor[HTML]{E6FFE6}(19, 19, 32) & \cellcolor[HTML]{E6FFE6}(18, 18, 37) & (48, 48, 48) \\ \hline
    7 & 72 & (34, 34, 71) & (28, 28, 55)                         & \cellcolor[HTML]{E6FFE6}(0, 0, 72) \\ \hline
    8   & 48 & \cellcolor[HTML]{E6FFE6}(22, 22, 37) & \cellcolor[HTML]{E6FFE6}(18, 18, 37) & (48, 48, 48) \\ \hline
    9 & 96 & \cellcolor[HTML]{E6FFE6}(25, 25, 61) & (37, 37, 74) & \cellcolor[HTML]{E6FFE6}(0, 0, 96) \\ \hline
    10 & 132 & \cellcolor[HTML]{E6FFE6}(43, 43, 81) & (51, 51, 102) & \cellcolor[HTML]{E6FFE6}(0, 0, 0)  \\ \hline
    Total & 780  & (300, 300, 600)  & (300, 300, 600) & (300, 300, 600) \\ \hline
    \end{tabular}
\end{table}

\begin{figure}[t]
    \centering
    \begin{subfigure}[t]{\linewidth}
        \centering
        \includegraphics[width=0.8\linewidth]{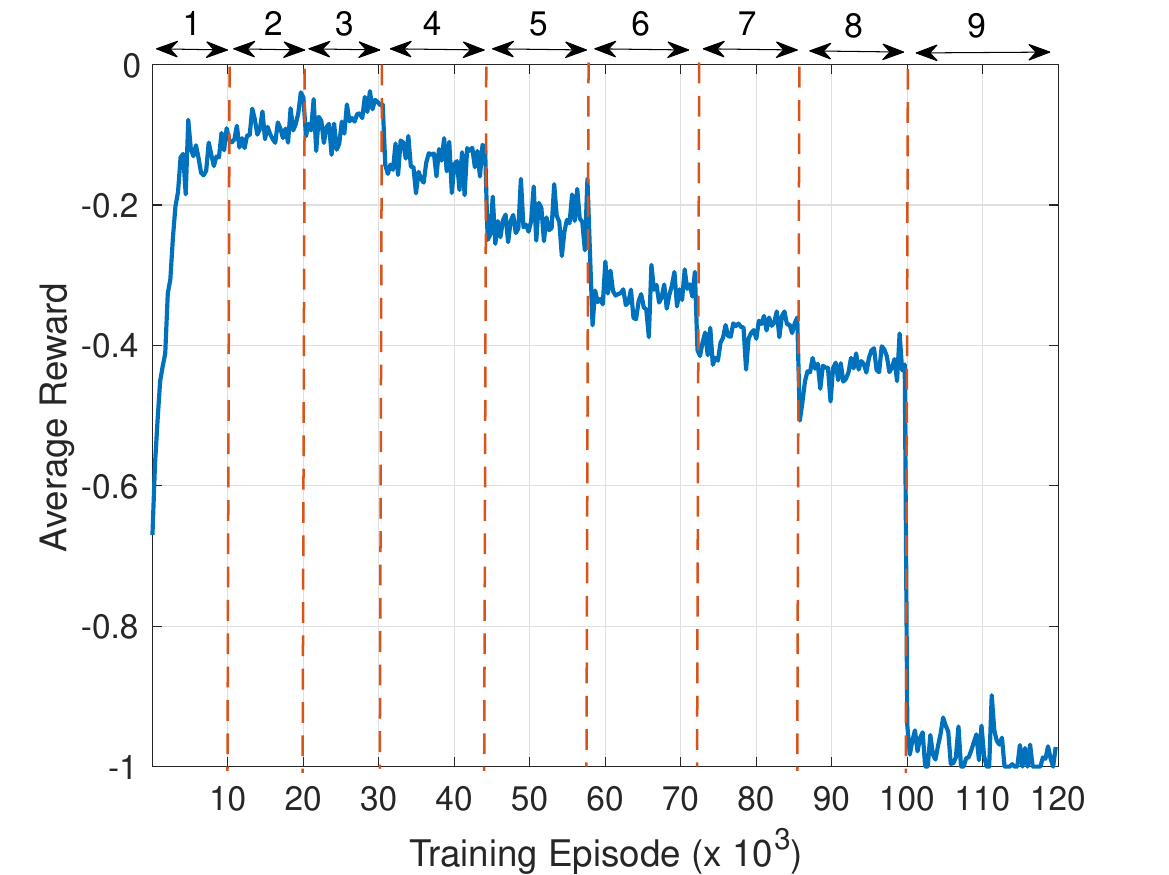}
        \caption{}
        \label{fig:RewadShowMorePackeImpact}
    \end{subfigure} 
    \vfill
    \begin{subfigure}[t]{\linewidth}
        \centering
        \includegraphics[width=0.8\linewidth]{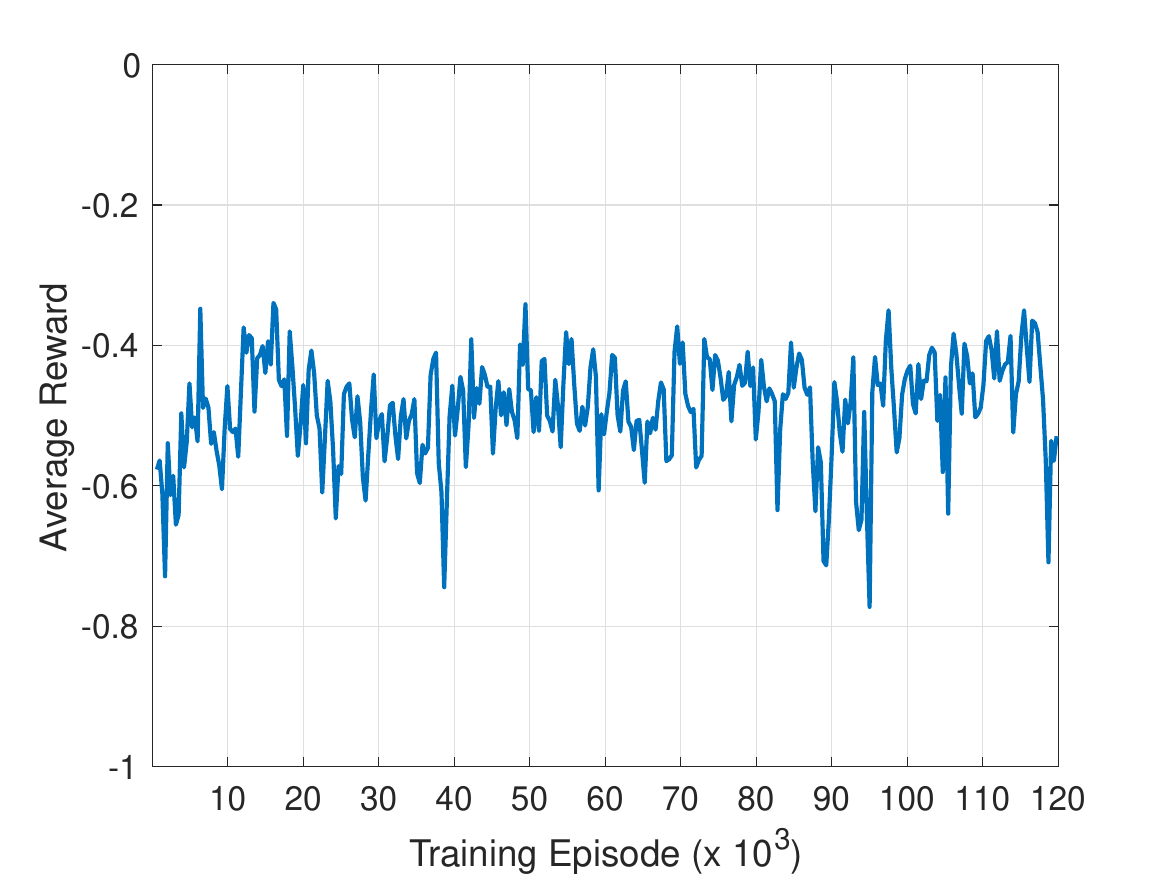}
        \caption{}
        \label{fig:Reward_before_modification}
    \end{subfigure}
    \caption{The learning behavior of Cyrus+. (a) Reward function under ACL with incresing number of URLLC packet arrivals.
    (b) Reward function when ACL is not employed.}
    \label{fig:RewardComparison}
\end{figure}

\noindent \textbf{Convergence of Learning} \ \ \ 
To evaluate the learning behavior of Cyrus+, we examine the reward function over TTIs.  
Figure~\ref{fig:RewardComparison}(a) shows the average reward vs. TTIs when the number of URLLC arrivals in a TTI varies from 1 to 9.  
Each point on the reward curve represents the averaged reward over the last $256$ TTIs.
Under ACL, for each URLLC arrival rate, we use a fixed eMBB window size.
As URLLC arrival rate increases, 
the eMBB window size is gradually reduced (see  Section~\ref{subsec:setting_pi_0_in_non_RT_ric}).
For a given eMBB window size, we repeat training for a number of episodes. 
Specifically, the window size and the number of episodes for that window size when URLLC packet arrival increases from 1 to 9 are: 
(50, $10^4$), 
(40, $10^4$), 
(30, $ 10^4$), 
(20, $1.4 \times 10^4$), 
(10, $1.4 \times 10^4$), 
(5, $1.4 \times 10^4$), 
(2, $1.4 \times 10^4$),
(1, $1.4 \times 10^4$), and 
(1, $1.4 \times 10^4$).
As shown in Fig~\ref{fig:RewardComparison}(a), the average reward successfully stabilizes at each setting (URLLC arrival rate).
When the number of URLLC packets reaches 8, the eMBB repetition window is reduced to 1, reflecting real-world operating conditions.
When the number of URLLC packet arrival in a TTI increase to 9, the average reward function drops down close to $-1$, indicating none of the eMBB packets can be decoded successfully.
That is, the URLLC packet arrival rate has exceeded the limit for multiplexing between eMBB and URLLC. 
Based on the above observations of the reward function, we store the NN parameters a the end of $10^5$ episodes and use them for $f_a$.

To demonstrate the significance of incorporating ACL in the learning process, 
Fig.~\ref{fig:RewardComparison}(b) presents the learning behavior of Cyrus+ after removing ACL.
In this case, training is conducted under real-world conditions, where the eMBB window size is one TTI and the number of URLLC packet arrivals varies randomly between 1 and 8 in each TTI. 
As shown in Fig.~\ref{fig:RewardComparison}(b), 
the average reward function 
fluctuates widely and fails to converge even after \(1.2\times 10^5\) TTIs.

\begin{figure}[t]
    \centering
    \includegraphics[width=0.8\linewidth]{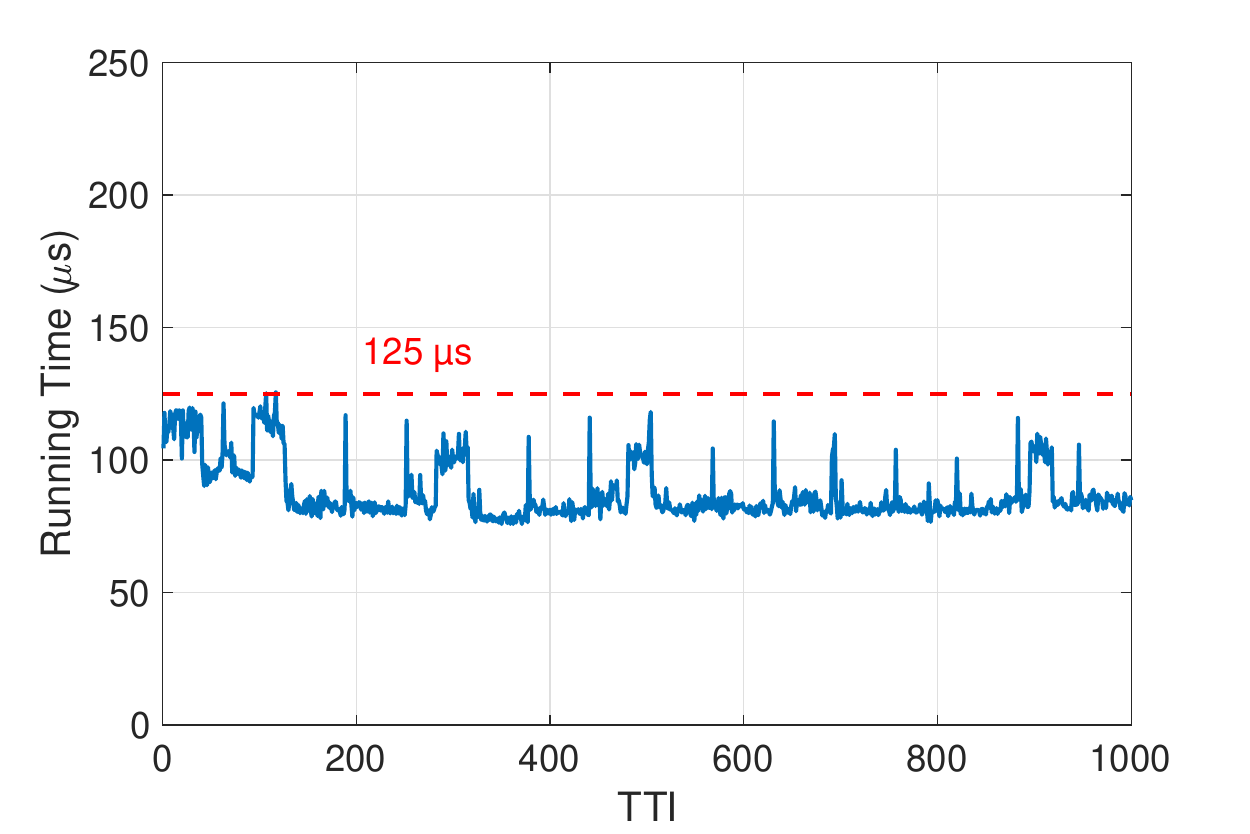}
    \caption{Computation time of Cyrus+ in each TTI for the RT O-DU.}
    \label{fig:Timing}
\end{figure}

\noindent \textbf{Timing Performance} \ \ \  
We now evaluate the timing performance of different control loops in Cyrus+.  
\begin{itemize} 

\item RT O-DU: The performance of this control loop is assessed by measuring the time elapsed from when the scheduling vector $\bm{s}(t)$ is received at the O-DU until the codebook $\bm{B}^*(t)$ is produced. 
Figure~\ref{fig:Timing} presents this running time for each TTI over 1,000 TTIs.
For all 1,000 TTIs, the running time for the RT O-DU is under 125 $\mu$s, which meets the timing requirement for 5G NR numerology 3. 
The average running time over 1,000 TTIs is $94.4$ $\mu$s.

\item Near-RT RIC: 
The performance of this control loop is assessed by measuring the time elapsed from the moment the critics fetch samples from the input buffer until the policy network $f_a$ is updated.
The average time measured over 1,000 training updates is $28.43$ \(m\)s.
This is well within the acceptable range for the Near-RT RIC in O-RAN, which spans from $10$ \(ms\) to $1 s$.

\item Non-RT RIC:  In the Non-RT RIC, the time to train the initial policy network $f_0$'s NN parameters takes 21 hours. 
Given that there is no specific time ceiling for the Non-RT RIC, we do not pursue any techniques to speed up the process. 
We refer readers to \cite{stooke2019:arXiv:acceleratedmethodsdeepreinforcement} if there is an interest to cut down this training time in the Non-RT RIC.

\end{itemize} 

\subsection{Varying System Parameters}
\label{subsec:varying_parameters}

We now investigate the behavior of Cyrus+ under varying system parameters.   

\noindent \textbf{Varying aggregate URLLC packet arrival rate} \ \ \
The aggregate URLLC arrival rate is determined by the number of active URLLC UEs and the probability of packet generation rate by each URLLC UE.
To vary the aggregate URLLC arrival rate, we can vary either one, which is equivalent.  
In this study, we fix the number of URLLC UEs at $12$ and vary the packet arrival probability $p$ in each mini-slot at each URLLC UE.
All other system parameters are the same as those in the case study. 

\begin{figure}[t]
    \centering
    \includegraphics[width=0.8\linewidth]{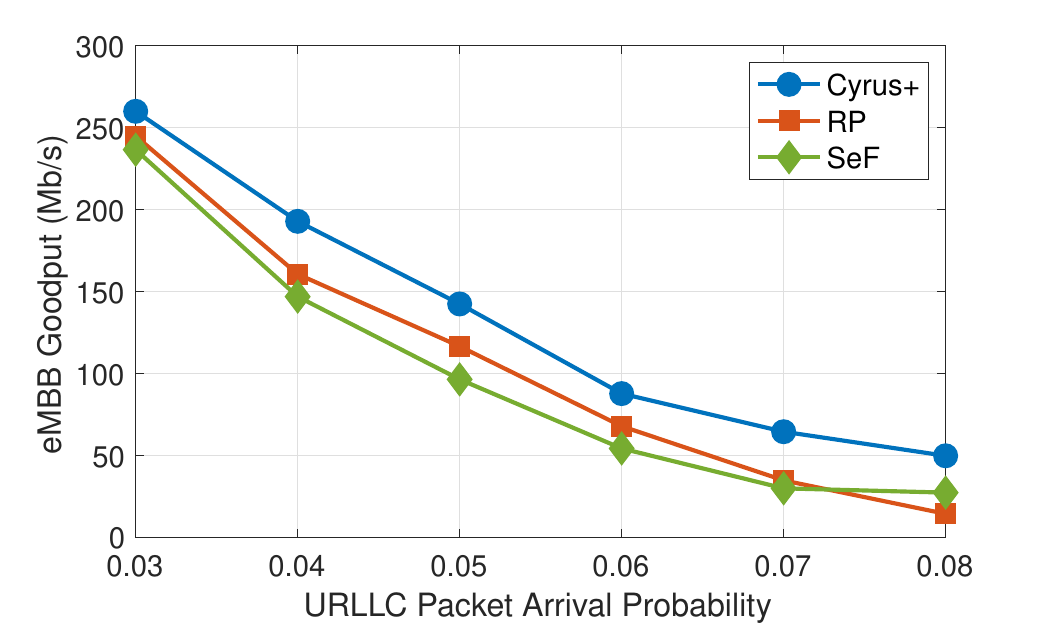}
    \caption{eMBB goodput under varying URLLC packet arrival rate $p$ in each mini-slot at each URLLC UE.}
    \label{fig:Goodput_FixURLLC_300SCs_10eMBBs}
\end{figure}

Figure~\ref{fig:Goodput_FixURLLC_300SCs_10eMBBs} shows the goodput performance of Cyrus+, as well as the other two benchmark schemes. 
As expected, the goodput decreases as $p$ increases across all schemes. 
Among the three schemes, Cyrus+ consistently achieves the best performance.
When the URLLC packet generation probability $p$ is low (e.g., $p=0.03$), RP and SeF exhibit performance similar to Cyrus+.
However, Cyrus+ performs noticeably better than RP and SeF as $p$ increases.  
When $p = 0.08$ (or about 1 URLLC packet per mini-slot on average or 7 URLLC packets per TTI), the goodput for Cyrus+ is 49.852 Mb/s, comparing to 14.430 Mb/s and 27.384 Mb/s for RP and SeF, respectively.

\begin{figure}[t]
    \centering
    \includegraphics[width=0.8\linewidth]{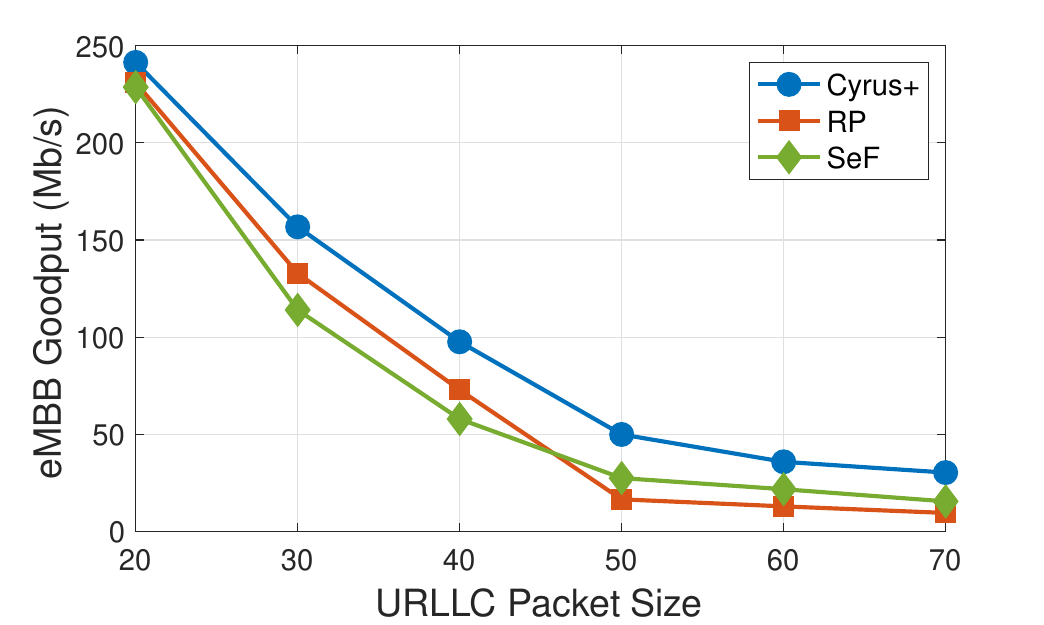}
    \caption{eMBB goodput under varying URLLC packet size.}
    \label{fig:Goodput_FixURLLC_192SCs_10eMBB}
\end{figure}

\noindent \textbf{Varying URLLC packet size} \ \ \
We now examine Cyrus+'s performance under varying URLLC packet size. 
All other system parameters remain 
the same as those in the case study.
Figure~\ref{fig:Goodput_FixURLLC_192SCs_10eMBB} show the results from our simulations.
As the URLLC packet size increases, the goodput decreases for Cyrus+, as well as the other two schemes.
But Cyrus+ achieves better goodput  performance than the other two schemes across all URLLC packet sizes.
When the URLLC packet size is 70 bytes (or $L = 420$ SCs for each URLLC packet), the goodput for the Cyrus+ is 30.221 Mb/s, compared to 9.428 Mb/s and 15.376 Mb/s for RP and SeF, respectively.

\begin{figure}[t]
    \centering
    \includegraphics[width=0.8\linewidth]{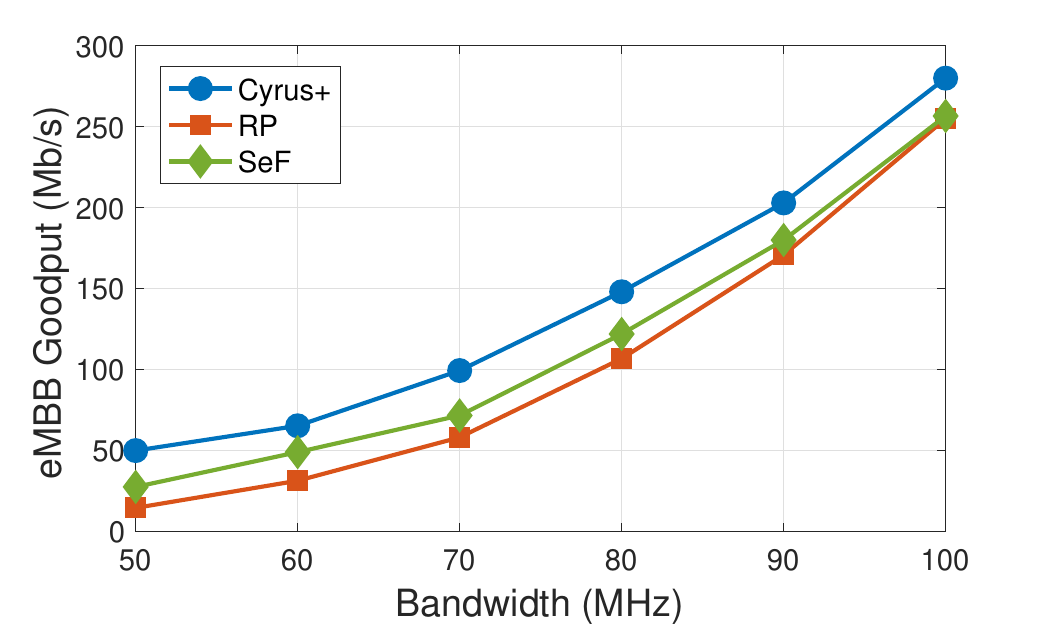}
    \caption{eMBB goodput under varying channel bandwidth.}
    \label{fig:GoodputVaryingBandwidth}
\end{figure}

\noindent \textbf{Varying available bandwidth} \ \ \
We now study Cyrus+'s performance under varying channel bandwidth. 
All other system parameters remain the same as those in the case study.
Figure~\ref{fig:GoodputVaryingBandwidth} presents the results.
As expected, as the channel bandwidth increases, the eMBB goodput in the cell increases.
Once again, Cyrus+ consistently outperforms both RP and SeF.
When the system bandwidth is 50 MHz, 
the goodput for the Cyrus+ is 49.852 Mb/s, compared to 14.430 Mb/s and 27.384 Mb/s for the RP and SeF, respectively.

\begin{figure}[t]
    \centering
    \begin{subfigure}[b]{0.8\linewidth}
        \centering
        \includegraphics[width=\linewidth]{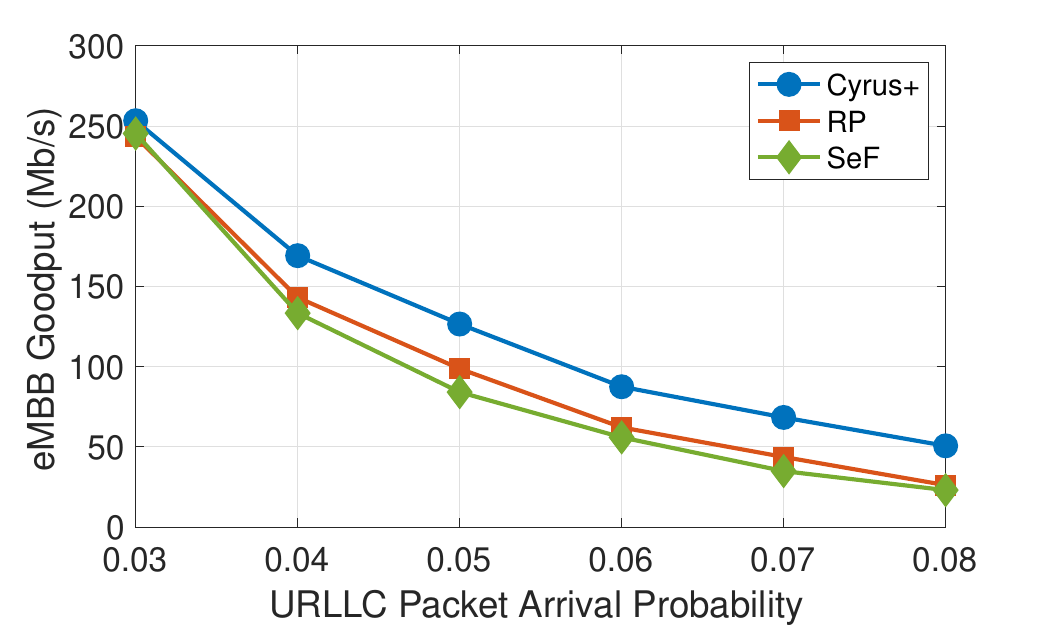}
        \caption{}
    \end{subfigure}
    
    \begin{subfigure}[b]{0.8\linewidth}
        \centering
        \includegraphics[width=\linewidth]{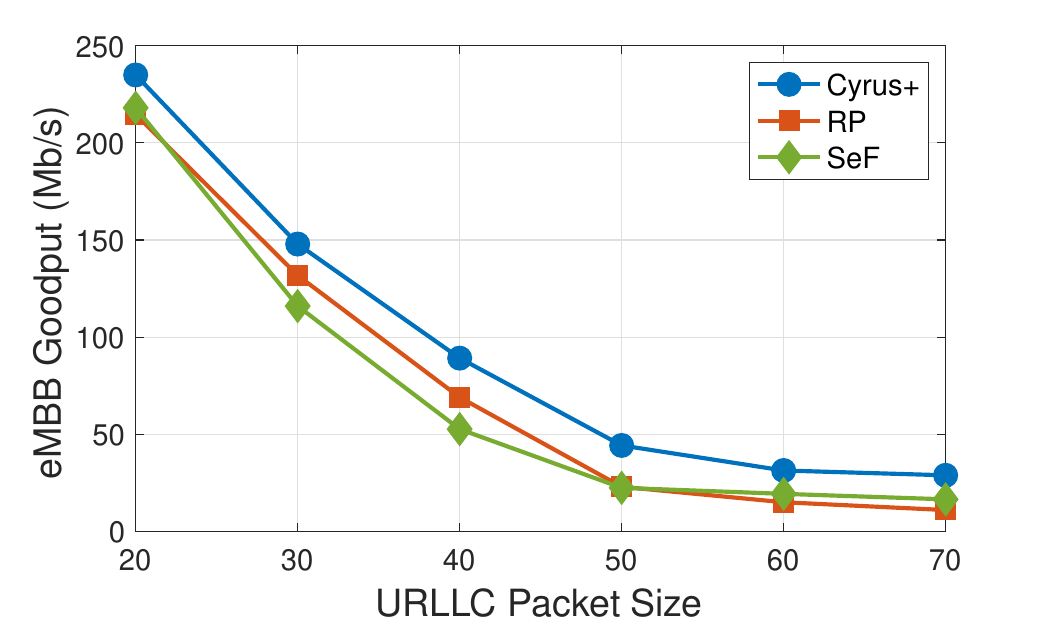}
        \caption{}
    \end{subfigure}
    
    \begin{subfigure}[b]{0.8\linewidth}
        \centering
        \includegraphics[width=\linewidth]{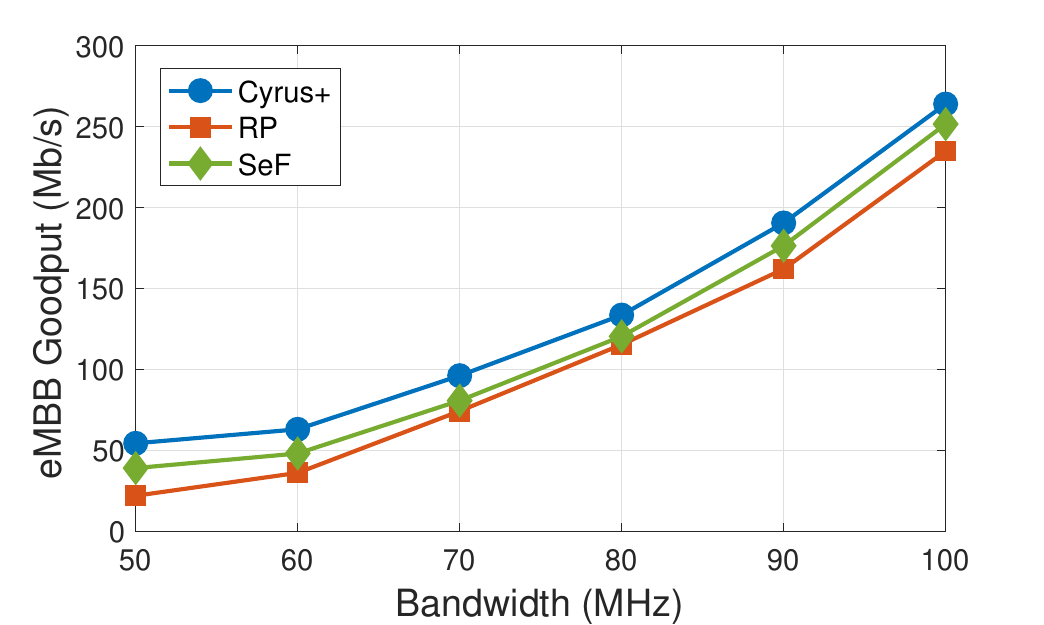}
        \caption{}
    \end{subfigure}
    
    \caption{eMBB goodput in a CDL-D channel under varying system parameters: 
    (a) URLLC packet arrival probabilities, (b) packet sizes, and (c) channel bandwidth.}
    \label{fig:Goodput_CDLD_Channel}
\end{figure}

\noindent \textbf{Varying channel model} \ \ \
All the above experimental results are conducted under the CDL-C channel model, which captures propagation behavior of a rich NLOS urban small cell environment.
To examine the robustness of Cyrus+ in other channel models, we consider CDL-D, which represents an urban macrocell. 
CDL-D incorporates a dominant LOS component along with Laplacian-distributed multipath components. 
Under the CDL-D channel model, we repeat all the above experiments for Cyrus+, as well as RP and SeF. 
Figure~\ref{fig:Goodput_CDLD_Channel} presents the results under varying aggregate URLLC packet arrival rate, varying URLLC packet size, and varying channel bandwidth.
The results are consistent to those under the CDL-C channel model, indicating that Cyrus+ is robust to different channel models.

\section{Conclusions}

In this paper, we investigated the problem of puncturing URLLC traffic onto eMBB transmissions on the same air interface.
We presented Cyrus+---a DRL-based solution for URLLC puncturing tailored for the O-RAN architecture. 
To have a precise measure of the impact of URLLC puncturing on eMBB, we employed goodput (based on the feedback of decoding outcome at each receiver), rather than estimated raw throughput (as in most of the prior efforts). 
At the core of Cyrus+ is a SAC-based method, whose dual-critic structure and entropy regularization enable stable training and high sample efficiency in high-dimensional action spaces.
In the RT O-DU, we proposed multiple parallel actors that concurrently process all possible URLLC arrivals to generate a puncturing codebook.
We introduced a feasibility enforcer in each branch that modifies any raw policy that violates the resource allocation constraints into a feasible solution.
In the Near-RT RIC, Cyrus+ customizes the critics' NN parameter update mechanism and the input buffer to match our design for the RT O-DU component.
In the Non-RT RIC, Cyrus+ employs ACL to stabilize training so as to ensure successful convergence of the DRL agent.
To validate Cyrus+'s performance, we developed a standard-compliant 5G NR simulator.
Experimental results demonstrated that Cyrus+ significantly outperforms two benchmark algorithms and meets the stringent timing requirement of 5G NR (numerology 3).


\bibliographystyle{IEEEtran} 
\bibliography{references}

\begin{thebibliography}{10}
\providecommand{\url}[1]{#1}
\csname url@samestyle\endcsname
\providecommand{\newblock}{\relax}
\providecommand{\bibinfo}[2]{#2}
\providecommand{\BIBentrySTDinterwordspacing}{\spaceskip=0pt\relax}
\providecommand{\BIBentryALTinterwordstretchfactor}{4}
\providecommand{\BIBentryALTinterwordspacing}{\spaceskip=\fontdimen2\font plus
\BIBentryALTinterwordstretchfactor\fontdimen3\font minus \fontdimen4\font\relax}
\providecommand{\BIBforeignlanguage}[2]{{%
\expandafter\ifx\csname l@#1\endcsname\relax
\typeout{** WARNING: IEEEtran.bst: No hyphenation pattern has been}%
\typeout{** loaded for the language `#1'. Using the pattern for}%
\typeout{** the default language instead.}%
\else
\language=\csname l@#1\endcsname
\fi
#2}}
\providecommand{\BIBdecl}{\relax}
\BIBdecl

\bibitem{Ghoreishi24:ICCCN:Cyrus}
E.~Ghoreishi, B.~Abolhassani, Y.~Huang, S.~Acharya, W.~Lou, and Y.~T. Hou, ``{Cyrus: A DRL-based Puncturing Solution to URLLC/eMBB Multiplexing in O-RAN},'' in \emph{Proc. 33rd International Conference on Computer Communications and Networks (ICCCN)}, pp. 1--9, July 29--31, 2024, {H}awaii, {USA}.

\bibitem{3gpp:tr38300:NROveralDiscription}
{3GPP}, \emph{{Service Requirements for the 5G System (3GPP TS 22.261, Version 19.6.0, Release 19, April 2024)}}, {A}vailable: \url{https://portal.3gpp.org/desktopmodules/Specifications/SpecificationDetails.aspx?specificationId=3107} (Last accessed: April 2024).

\bibitem{3gpp:tr38824:urllc}
3GPP, \emph{{Study on Physical Layer Enhancements for NR Ultra-Reliable and Low Latency Case (URLLC) (3GPP TR 38.824, Version 16.0.0, Release 16, March 2019)}}, {Available}: \url{https://portal.3gpp.org/desktopmodules/Specifications/SpecificationDetails.aspx?specificationId=3498} (Last accessed: April 5, 2024).

\bibitem{3gpp:meeting88:Puncturing}
{{3GPP}}, \emph{{Final Report of 3GPP TSG RAN WG1 Meeting \#88 (Version 1.0.0, February 2017)}}, {A}vailable: \url{https://www.3gpp.org/ftp/TSG_RAN/WG1_RL1/TSGR1_88/Report/} (Last accessed: April 3, 2024).

\bibitem{3gpp:tr38211:PhysicalChannelsModulation}
{3GPP}, \emph{{NR; Physical channels and modulation (3GPP TS 38.211, Version 18.4.0, Release 18, September 2024)}}, {A}vailable: \url{https://portal.3gpp.org/desktopmodules/Specifications/SpecificationDetails.aspx?specificationId=3214} (Last accessed: October 1, 2024).

\bibitem{Anand18:INFOCOM:Joint_URLLC_eMBB}
{A. Anand, G. De Veciana, and S. Shakkottai}, ``{Joint Scheduling of URLLC and eMBB Traffic in 5G Wireless Networks},'' in \emph{{Proc. IEEE INFOCOM}}, pp. 1970--1978, April 16--19, 2018, {H}onolulu, HI, USA.

\bibitem{O-RANAlliance2023:site:ORAN}
{O-RAN Alliance}, \emph{{O-RAN Architecture Description}}, {A}vailable: \url{https://orandownloadsweb.azurewebsites.net/specifications} (Last accessed: January 5, 2024).

\bibitem{haarnoja18:ICML:SAC}
T.~Haarnoja, A.~Zhou, P.~Abbeel, and S.~Levine, ``{Soft Actor-Critic: Off-Policy Maximum Entropy Deep Reinforcement Learning with a Stochastic Actor},'' in \emph{Proc. ICML}, pp. 1861--1870, July 10--15, 2018, {S}tockholm Sweden.

\bibitem{Alqwider24:transaction:URLLCpuncturing_eMBBscheduling}
W.~Alqwider, A.~S. Abdalla, T.~F. Rahman, and V.~Marojevic, ``{Intelligent Dynamic Resource Allocation and Puncturing for Next-Generation Wireless Networks},'' \emph{{IEEE Internet of Things Journal}}, vol. 11, no. 19, pp. 31438--31452, 2024.

\bibitem{Sohaib23:IEEEAccess:DRLapproximationDecoding}
R.~M. Sohaib, O.~Onireti, Y.~Sambo, R.~Swash, S.~Ansari, and M.~A. Imran, ``{Intelligent Resource Management for eMBB and URLLC in 5G and Beyond Wireless Networks},'' \emph{IEEE Access}, vol. 11, pp. 65205-65221, 2023.

\bibitem{Fabio:GlobCom:DRL_threshold}
F.~Saggese, L.~Pasqualini, M.~Moretti, and A.~Abrardo, ``{Deep Reinforcement Learning for URLLC data management on top of scheduled eMBB traffic},'' in \emph{Proc. IEEE GLOBECOM}, pp. 1--6, December 7--11, 2021, {M}adrid, {S}pain.

\bibitem{Huang20:IOT:DEMUX}
Y.~Huang, S.~Li, C.~Li, Y.~T. Hou, and W.~Lou, ``{A Deep-Reinforcement-Learning-Based Approach to Dynamic eMBB/URLLC Multiplexing in 5G NR},'' \emph{{IEEE Internet of Things Journal}}, vol.~7, no.~7, pp. 6439--6456, 2020.

\bibitem{Bairagi:TC:Coexistence}
A.~K. Bairagi, M.~S. Munir, M.~Alsenwi, N.~H. Tran, S.~S. Alshamrani, M.~Masud, Z.~Han, and C.~S. Hong, ``{Coexistence Mechanism Between eMBB and uRLLC in 5G Wireless Networks},'' \emph{IEEE Transactions on Communications}, vol.~69, no.~3, pp. 1736--1749, 2021.

\bibitem{Alsenwi19:IEEEComLet:puncturing}
M.~Alsenwi, N.~H. Tran, M.~Bennis, A.~Kumar~Bairagi, and C.~S. Hong, ``{eMBB-URLLC Resource Slicing: A Risk-Sensitive Approach},'' \emph{IEEE Communications Letters}, vol.~23, no.~4, pp. 740--743, 2019.

\bibitem{Almekhlafi21:ICC:threshold}
M.~Almekhlafi, M.~A. Arfaoui, C.~Assi, and A.~Ghrayeb, ``{Joint Resource and Power Allocation for URLLC-eMBB Traffics Multiplexing in 6G Wireless Networks},'' in \emph{{Proc. IEEE ICC}}, pp. 1-6, June 14--23, 2021, {M}ontreal Canada.

\bibitem{Almekhlafi21:transaction:puncturing_probability}
M.~Almekhlafi, M.~Chraiti, M.~A. Arfaoui, C.~Assi, A.~Ghrayeb, and A.~Alloum, ``{A Downlink Puncturing Scheme for Simultaneous Transmission of URLLC and eMBB Traffic by Exploiting Data Similarity},'' \emph{{IEEE Transactions on Vehicular Technology}}, vol. 70, no. 12, pp. 13087-13100, 2021.

\bibitem{Gallager1963:IEEETrans:LDPC}
R.~G. Gallager, \emph{Low-Density Parity-Check Codes}, ser. Monograph No. RLE-TR-156.\hskip 1em plus 0.5em minus 0.4em\relax Cambridge, MA: MIT Press, 1963.

\bibitem{3gpp23:tr38201:PHY}
{3GPP}, \emph{{NR; Physical layer; General description (3GPP TS 38.201, Version 18.0.0, Release 18, Sep. 2023)}}, {A}vailable: \url{https://portal.3gpp.org/desktopmodules/Specifications/SpecificationDetails.aspx?specificationId=3211} (Last accessed: Apr. 8, 2024).

\bibitem{Rybin14:IEEE:errorcorection}
P.~Rybin, ``{On the error-correcting capabilities of low-complexity decoded irregular LDPC codes},'' in \emph{Proc. IEEE International Symposium on Information Theory}, pp. 3165--3169, Jun 30-- July 5, 2014, {H}onolulu, {H}awai, {USA}.

\bibitem{Richardson18:BLER:Numerically}
T.~Richardson and S.~Kudekar, ``{Design of Low-Density Parity Check Codes for 5G New Radio},'' \emph{IEEE Communications Magazine}, vol.~56, no.~3, pp. 28--34, 2018.

\bibitem{Schulman:arXiv:PPO}
J.~Schulman, F.~Wolski, P.~Dhariwal, A.~Radford, and O.~Klimov, ``{Proximal Policy Optimization Algorithms},'' arXiv:1707.06347, 2017.

\bibitem{Haijun20:SpectrumSharing:PPO}
H.~Zhang, N.~Yang, W.~Huangfu, K.~Long, and V.~C.~M. Leung, ``{Power Control Based on Deep Reinforcement Learning for Spectrum Sharing},'' \emph{{IEEE Transactions on Wireless Communications}}, vol.~19, no.~6, pp. 4209--4219, 2020.

\bibitem{Timothy19:arXiv:DDPG}
T.~P. Lillicrap, J.~J. Hunt, A.~Pritzel, N.~Heess, T.~Erez, Y.~Tassa, D.~Silver, and D.~Wierstra, ``Continuous control with deep reinforcement learning,'' arXiv:1509.02971, 2019.

\bibitem{Xuemin20:DDPG:resourceManagment}
H.~Peng and X.~Shen, ``{Deep Reinforcement Learning Based Resource Management for Multi-Access Edge Computing in Vehicular Networks},'' \emph{{IEEE Transactions on Network Science and Engineering}}, vol.~7, no.~4, pp. 2416--2428, 2020.

\bibitem{Fujimoto18:Arkiv:TD3}
S.~Fujimoto, H.~van Hoof, and D.~Meger, ``{Addressing Function Approximation Error in Actor-Critic Methods},'' in \emph{Proc. ICML}, pp. 1582--1591, July 10--15, 2018, {S}tockholm Sweden.

\bibitem{sutton2018:Book:RL:AnIntroduction}
R.~S. Sutton and A.~G. Barto, \emph{Reinforcement Learning: An Introduction}, 2nd~ed.\hskip 1em plus 0.5em minus 0.4em\relax Cambridge, MA, USA: MIT Press, 2018.

\bibitem{kullback1951:JSTOR:KL}
S.~Kullback and R.~A. Leibler, ``{On Information and Sufficiency},'' \emph{The Annals of Mathematical Statistics}, vol.~22, no.~1, pp. 79--86, 1951.

\bibitem{duchi2008:book:waterfilling}
J.~Duchi, S.~Shalev-Shwartz, Y.~Singer, and T.~Chandra, ``{Efficient projections onto the l 1-ball for learning in high dimensions},'' in \emph{Proc. ICML}, pp. 272--279, July 5--9, 2008, {H}elsinki, {F}inland.

\bibitem{darroch1972:JSTOR:KLSolver}
J.~N. Darroch and D.~Ratcliff, ``{Generalized Iterative Scaling for Log-Linear Models},'' \emph{The Annals of Mathematical Statistics}, vol. 43, no. 5, pp. 1470--1480, 1972.

\bibitem{Balinski2010:intiger}
M.~L. Balinski and H.~P. Young, \emph{{Fair Representation: Meeting the Ideal of One Man, One Vote}}.\hskip 1em plus 0.5em minus 0.4em\relax Washington, DC, USA: Brookings Institution Press, 2001.

\bibitem{Yoshua:Curriculmlearning:ICML}
Y.~Bengio, J.~Louradour, R.~Collobert, and J.~Weston, ``{Curriculum learning},'' in \emph{Proc. ICML}, pp. 41--48, June 14--18, 2009, {Q}uebec, Canada.

\bibitem{Portelas20:AutomaticCL:survay}
R.~Portelas, C.~Colas, L.~Weng, K.~Hofmann, and P.-Y. Oudeyer, ``{Automatic Curriculum Learning For Deep RL: A Short Survey},'' in \emph{Proc. 29th International Joint Conference on Artificial Intelligence}, pp. 1--7, January 7--15, 2021, {Y}okohama, {J}apan.

\bibitem{Carlos17:ReverseCurriculum}
C.~Florensa, D.~Held, M.~Wulfmeier, and P.~Abbeel, ``{Reverse Curriculum Generation for Reinforcement Learning},'' in \emph{Proc. Conference on Robot Learning}, pp. 482--495, November 13--15, 2017, {C}alifornia, {USA}.

\bibitem{MathWorks2023:online:5GToolbox}
{MathWorks}, ``{5G Toolbox Documentation},'' 2023, {A}vailable: \url{https://www.mathworks.com/help/5g/} (Last accessed: April 5, 2024).

\bibitem{abadi2016:arxiv:tensorflow}
{M. Abadi et al.}, ``{TensorFlow: Large-Scale Machine Learning on Heterogeneous Distributed Systems},'' 2016, arXiv:1603.04467.

\bibitem{Kelly98:JonOprational:PFScheduler}
{F. Kelly, A. Maulloo, and D. Tan}, ``{Rate Control for Communication Networks: Shadow Prices, Proportional Fairness and Stability},'' \emph{Journal of the Operational Research Society}, vol.~49, no.~3, pp. 237--252, 1998.

\bibitem{NVIDIA2024:online:cuDNN}
{NVIDIA}, ``{cuDNN Developer Guide V9.1.0},'' Available: \url{https://docs.nvidia.com/deeplearning/cudnn/latest/developer/overview.html}, 2024, {(Last accessed: April 12, 2024)}.

\bibitem{kingma13:arXiv:ParametrizeTrick}
D.~P. Kingma and M.~Welling, ``{Auto-Encoding Variational Bayes},'' arXiv:1312.6114, 2022.

\bibitem{ha2003:Book:LDPCPunctruing}
J.~Ha, ``{Low-Density Parity-Check Codes with Erasures and Puncturing},'' {P}h.D. dissertation, Georgia Institute of Technology, November 2003, {A}tlanta, USA.

\bibitem{Nair10:ICML:ReLU}
V.~Nair and G.~E. Hinton, ``{Rectified Linear Units Improve Restricted Boltzmann Machines},'' in \emph{{Proc. ICML}}, pp. 807--814, June 21--14, 2010, {H}aifa, Israel.

\bibitem{kingma2017:Arxiv:adam}
D.~P. Kingma and J.~Ba, ``{Adam: A Method for Stochastic Optimization},'' arXiv:1412.6980, 2016.

\bibitem{stooke2019:arXiv:acceleratedmethodsdeepreinforcement}
A.~Stooke and P.~Abbeel, ``{Accelerated Methods for Deep Reinforcement Learning},'' arXiv:1803.02811, 2019.

\end{thebibliography}

\end{document}